\documentclass[11pt]{article}

\usepackage{epsfig,multicol,bbm}
\usepackage{longtable}
\usepackage{amsthm,latexsym,multibox,amssymb,amsfonts,array}
\usepackage[centertags,intlimits]{amsmath}
\usepackage{eufrak}
\usepackage[mathscr]{eucal}
\usepackage{graphicx}
\newcommand{\beq}{\begin{equation}}
\newcommand{\eeq}{\end{equation}}
\newcommand{\eqnlab}[1]{\label{eqn:#1}}
\newcommand{\Eqnref}[1]{Eq.~(\ref{eqn:#1})}
\newcommand{\beqa}{\begin{eqnarray}}
\newcommand{\eqa}{\end{eqnarray}}
\newcommand{\hs}{\hspace{0.1 cm}}

\newcommand{\al}{\alpha}
\newcommand{\be}{\beta}

\newcommand{\si}{\sigma}

\newcommand{\f}{\frac}
\newcommand{\mc}{\mathcal}

\newcommand{\nn}{\nonumber}

\newcommand{\mf}{\mathfrak}

\newcommand{\ph}{\phantom}

\newcommand{\mbb}{\mathbb}

\newcommand{\cg}{\mathbb{G}_{\mf{C}}}
\newcommand{\tbf}{\textbf}
\long\def\symbolfootnote[#1]#2{\begingroup%
\def\thefootnote{\fnsymbol{footnote}}\footnote[#1]{#2}\endgroup}
\begin{document}

\begin{flushright} ULB-TH/06-26
\end{flushright}
\vspace{.7cm}
\begin{centering}

{\Large{\textbf{A Special Class of Rank $10$ and $11$ Coxeter Groups\symbolfootnote[2]{This version contains an Erratum, correcting a mistake regarding the number of symmetric rank $11$ Coxeter graphs with incidence index $4$. The bulk of the paper is unchanged, except for scattered comments drawing the reader's attention to where the Erratum applies.}}}} \\

\vspace{1.1cm}
\rule[0.1in]{14cm}{0.5mm} \\
\vspace{.9cm}
 Marc Henneaux ${}^{\spadesuit \diamondsuit}$, Mauricio Leston ${}^{\clubsuit \triangle}$,
 Daniel Persson ${}^{\spadesuit}$ and Philippe Spindel ${}^{\heartsuit}$\\
\vspace*{1cm} {\footnotesize
 ${}^{\spadesuit}$  Physique Th\'eorique et Math\'ematique,\\
 Universit\'e Libre de Bruxelles and International Solvay Institutes, \\
 Boulevard du Triomphe, ULB -- C.P. 231, B-1050 Bruxelles, Belgium  \\
 \vspace{.2cm}
 ${}^{\diamondsuit}$ Centro de Estudios Cient\'{\i}ficos
(CECS), Casilla 1469, Valdivia, Chile\\
 \vspace{.2cm}
${}^\triangle$ Theoretische Naturkunde, Vrije Universiteit Brussel and   \\
  International Solvay Institutes, Pleinlaan 2, B-1050 Brussels, Belgium\\
\vspace{.2cm}
${}^\clubsuit$   Instituto de Astronomica y Fisica del Espacio (IAFE), \\
  Casilla de Correo 67, Sucursal 28, 1428 Buenos Aires, Argentina\\
   \vspace{.2cm}
${}^{\heartsuit}$   Service de M\'ecanique et Gravitation,\\
  Universit\'e de Mons-Hainaut, Acad\'emie Wallonie-Bruxelles,\\
 Avenue du Champ de Mars 6, 7000 Mons, Belgium\\
 \vspace{.2cm}
E-mail: \textbf{henneaux@ulb.ac.be}, \textbf{mauricio@iafe.uba.ar},
\textbf{dpersson@ulb.ac.be}, \textbf{Philippe.Spindel@umh.ac.be}}

\vspace{1.5cm}
\begin{minipage}{.9\textwidth}
\abstract{In the course of investigating regular subalgebras of
$E_{10(10)}$ related to cosmological solutions of 11-dimensional
supergravity supporting an electric 4-form field, a class of rank
10 Coxeter subgroups of the Weyl group of $E_{10(10)}$ was
uncovered (hep-th/0606123). These Coxeter groups all share the
property that their Coxeter graphs have incidence index 3, i.e.
that each node is incident to three and only three single lines.
Furthermore, the Coxeter exponents are either 2 or 3, but never
$\infty$. We here go beyond subgroups of the Weyl group of
$E_{10(10)}$ and classify all rank 10 Coxeter graphs with these
properties. We find 21 distinct Coxeter groups of which 7 were
already described in hep-th/0606123. Moreover, we extend the
classification to the rank 11 case and we find 252 inequivalent
rank 11 Coxeter groups with incidence index 4, of which at least
28 can be regularly embedded into $E_{11(11)}$. }
\end{minipage}
\vspace{6cm}

\end{centering}
\thispagestyle{empty}
%\tableofcontents

\section{Introduction}

The study of hidden symmetries of supergravity theories has
revealed an intriguing connection with the theory of infinite
dimensional Kac-Moody algebras. This correspondence has been most
extensively explored from various directions in the context of
eleven dimensional supergravity. In particular, the theory
exhibits an exceptional coset symmetry
$\mc{E}_{d(d)}/\mc{K}(\mc{E}_{d(d)})$ when compactified on a
$d$-torus $T^{d}$ (see e.g.
\cite{Dualisation}\cite{Cosetsymmetries}). This symmetry combines
the non-perturbative S-duality and the perturbative T-duality of
string theory into the so-called U-duality group $\mc{E}_{d(d)}$.
It has been conjectured that the discrete subgroup
$\mc{E}_{d(d)}(\mbb{Z})$ lifts to a symmetry of the full string
theory \cite{HullTownsend}\cite{ObersPioline}.

The appearance of these symmetries is well established for
compactification down to $D=2+1$ dimensions where the U-duality
symmetry is described by the split form of the largest of the
exceptional Lie groups, namely $\mc{E}_{8(8)}$. It is furthermore
known that in $D=1+1$ dimensions there is an affine symmetry group
leaving the equations of motion invariant. This group is the analogue of
the Geroch group and has been identified with the affine extension
of $\mc{E}_{8(8)}$, denoted $\mc{E}_{9(9)}\equiv
\mc{E}_{8(8)}^{+}$. The affine symmetry is responsible for the
emergence of an integrable structure of $N=16$ supergravity
reduced to $2$ dimensions \cite{NicolaiIntegrability}\cite{NicolaiWarner}\cite{NicolaiSamtleben}.

Further compactification on $T^{10}$ and $T^{11}$ is much less
understood but it was conjectured long ago by Julia \cite{Julia}
that the chain of extended symmetries should somehow remain
unbroken, thus giving rise to $\mc{E}_{10(10)}$ and
$\mc{E}_{11(11)}$ as symmetry groups of eleven dimensional
supergravity appropriately compactified to $D=1$ and $D=0$
dimensions, respectively.

Subsequently, West showed \cite{Westhidden}\cite{West} that eleven
dimensional supergravity could be reformulated as a non-linear
realization based on a finite dimensional Lie algebra called
$G_{11}$ whose structure corresponded in part to the low-level structure of the
infinite dimensional Kac-Moody algebra
$E_{11(11)}=\text{Lie}\big[\mc{E}_{11(11)}\big]$. This led him to
conjecture that $E_{11(11)}$ should in fact be a symmetry of the
full (uncompactified) eleven dimensional supergravity, or even of
M-theory itself.

Another window into the Kac-Moody structure of M-theory was later
opened through the study of eleven dimensional supergravity close
to a spacelike singularity. It was found that the effective
dynamics at each spatial point could be mapped onto a piecewise
linear particle motion in an auxiliary space of Lorentzian
signature. The piecewise nature of the motion is due to
reflections against hyperplanes in the Lorentzian space and these
reflections form a Coxeter group which could be identified with
the Weyl group of $E_{10(10)}=\text{Lie}\big[\mc{E}_{10(10)}\big]$
\cite{Arithmetical}\cite{Damour:2002et}. This unexpected result
was taken to be a strong indication that  the
$E_{10(10)}$-symmetry should play a fundamental role in the
ultimate formulation of M-theory.

The relation between $E_{10(10)}$ and M-theory was pushed even
further in \cite{Damour:2002cu} where a Lagrangian based on the
coset space $\mc{E}_{10(10)}/\mc{K}(\mc{E}_{10(10)})$ was
explicitly constructed and whose dynamics reproduced the dynamics
of a certain regime of eleven dimensional supergravity. This
conjectural and somewhat mysterious relation between the dynamics
of M-theory and the geodesic flow on
$\mc{E}_{10(10)}/\mc{K}(\mc{E}_{10(10)})$ has up until now been
thoroughly tested and verified only within consistently truncated
versions of both theories.

By a \emph{consistent truncation} we mean a
truncation such that a solution to the truncated equations of
motion is also a solution to the equations of motion of the full
theory. A natural truncation from the algebraic point of view is the
\emph{level truncation}. The level provides a grading of
$E_{10(10)}$ and one may consistently
truncate the theory to any finite level by setting all higher
level ``covariant derivatives" to zero \cite{Damour:2002cu}.
Another useful type of truncation is that of a \emph{subgroup
truncation}. One restricts the dynamics on the coset space to an
appropriately chosen subgroup, say $\bar{\mc{G}}\subset
\mc{E}_{10(10)}$. The equations of motion imply that the evolution of
initial data in $\bar{\mc{G}}$ remain in $\bar{\mc{G}}$. As long
as we restrict to subgroups generated by fields that live within
the established region of compatibility with the supergravity
dynamics we know that solutions of the equations of motion for the
sigma model $\bar{\mc{G}}/\mc{K}(\bar{\mc{G}})$ also correspond to
exact solutions on the supergravity side.

In the cosmological regime one usually imposes conditions on the
metric and the $4$-form field strength in order to simplify the
dynamics. A class of electric solutions to $11$-dimensional
supergravity was found long ago in \cite{Demaret:1985js} by
considering a diagonal spatial metric and a diagonal energy
momentum tensor of the $4$-form. An interesting subclass of this
class of solutions can be encoded in so-called \emph{geometric
configurations} $(n_m,g_3)$. These consist of $n$ points and $g$
lines, drawn on the plane with the following rules; (i) each line
is incident to three points, (ii) each point is incident to $m$
lines and (iii) two points determine at most one line. One then
associates a non-vanishing component of the electric field,
$F_{0ijk}$, for each line in the configuration and one takes a
zero magnetic field. The third condition above encodes the
diagonality of the energy-momentum tensor of $F_{\mu\nu\rho\si}$.

Cosmological solutions to eleven-dimensional supergravity were
recently reinvestigated in \cite{Kleinschmidt:2005gz} from the
point of view of the proposed connection between M-theory and
$E_{10(10)}$. This was further pursued  by the present authors in
\cite{GeometricConf} where it was found that each geometric
configuration discussed in \cite{Demaret:1985js} corresponds in
fact to a regular subalgebra, $\bar{\mf{g}}$, of $E_{10(10)}$ and that
the solutions of the corresponding sigma model, i.e. truncated to
the relevant regular subalgebra, generalize the cosmological
solutions described in \cite{Demaret:1985js}. Some of these
solutions correspond to so-called S-brane solutions and it was
noted that the geometric configurations provide information about
the intersection rules between S-branes\footnote{See \cite{Argurio}\cite{Ohta} for the original analysis of intersection rules for $p$- and S-branes. See also \cite{Ivashchuk}\cite{Ivashchuk2} for related discussions in the context of hyperbolic Kac-Moody algebras.}. This result is in spirit
very similar to the analysis of \cite{IntersectingBranes}, where
brane solutions were analyzed in the context of very extended
Kac-Moody algebras.

The Dynkin diagram $\mbb{D}_{\bar{\mf{g}}}$ derived from a certain
configuration $(n_m,g_3)$ is the \emph{line-incidence diagram} of
$(n_m,g_3)$, meaning that each line in $(n_m,g_3)$ defines a node
in $\mbb{D}_{\bar{\mf{g}}}$ and two nodes in
$\mbb{D}_{\bar{\mf{g}}}$ are connected only if the corresponding
lines in $(n_m,g_3)$ are parallel. In particular, a set of six
rank $10$ Lorentzian subalgebras of $E_{10(10)}$ was uncovered in
this way from the configurations with $10$ points and $10$ lines,
$(n_m,g_3)=(10_3,10_3)$. As a consequence of the rules for
constructing the configurations, the Dynkin diagrams of these
algebras displayed a remarkable regularity: all nodes in the
diagrams are connected to three and only three other nodes, or, in
other words, each node in the Dynkin diagrams is incident to three
single lines.  {}Furthermore, the Cartan indices (off-diagonal
entries of the corresponding Cartan matrix, which is symmetric in
the present context) are equal to $0$ or $-1$ and do not take
lower negative values ($-2, -3 , \cdots$) even though these are in
principle allowed within $E_{10(10)}$ (as pointed out in the next
section).

Since none of these algebras are hyperbolic, the corresponding
gravitational dynamics is non-chaotic and, in the BKL-limit, the
solution settles asymptotically into a Kasner solution after
finitely many oscillations \cite{nonchaos,Demaret:1986ys}.

In the billiard analysis for generic solutions of 11-dimensional
supergravity, one only sees the Weyl group of $E_{10(10)}$ because
this is the group of reflections in the walls bounding the
billiard table, i.e. reflections in the simple roots of $E_{10(10)}$.
From this point of view the analysis of \cite{GeometricConf}
revealed a new class of rank $10$ Coxeter subgroups of the Weyl
group of $E_{10}$. Because of the fact that the Dynkin diagrams of the Kac-Moody
algebras derived from the configurations $(10_3,10_3)$ are all
connected and the corresponding Cartan matrices are symmetric, it follows that the
associated Coxeter exponents can only be $2$, $3$ or $\infty$ \cite{Humphreys}.
However, the value $\infty$ corresponding to the infinite dihedral
group does not arise in the particular subclass analyzed in
\cite{GeometricConf} since the only values of the Cartan integers
are $0$ or $-1$. The Coxeter graphs inherit in addition the
property of the Dynkin diagrams that the nodes are all incident to
exactly three lines.

In fact, it was shown in \cite{GeometricConf} that in all
geometric configurations, $(n_m,g_3)$, each line is parallel to a
number $k$ of different lines, where $k$ depends on the
configuration but not on the individual lines. This implies that
the Coxeter graphs derived from such configurations have the
property that each node, $P$, in the graph is connected to a
number $k$ of other nodes, such that $k$ depends on the
configuration but not on the node $P$ itself. Henceforth we will
refer to the number $k$ as the \emph{incidence index} and denote
it by $\mc{I}$.

The purpose of this short note is to go beyond Coxeter groups
determined by geometric configurations and classify all rank $10$
Coxeter groups with the following two properties:
\begin{itemize} \item the incidence index $\mc{I}$ is equal to $3$,
$\mc{I}=3$;
\item the Coxeter exponents $m_{ij}$ ($i \not=j$, see below) are
equal to $2$ or $3$.
\end{itemize} The classification is equivalent to the classification of
all symmetric rank $10$ Cartan matrices with off-diagonal terms
equal to $0$ or $-1$.

The classification is done by constructing the Cartan matrices
from scratch in a step-by-step procedure. In this way we find that
the rank $10$ Coxeter groups come in three distinct classes:
\begin{itemize}
\item $9$ Cartan matrices of Lorentzian signature, $6$ of which
correspond to geometric configurations.
\item $2$ degenerate cases whose Cartan matrices both have
vanishing determinants but with one negative eigenvalue each.
One of these cases can be derived from a geometric configuration.
\item $10$ Cartan matrices with positive determinants but
with signatures $(2\big|_{-},8\big|_{+})$.
\end{itemize}

The same classification scheme is further pursued for the case of
rank $11$ Coxeter groups with the following result\footnote{This list is incomplete; see the Errata added on page 19.}:

\begin{itemize}
\item 71 Cartan matrices of Lorentzian signature, 15 of which
correspond to geometric configurations and so can be embedded into
$E_{11(11)}$. \item 5 Cartan matrices with negative determinants,
all of which have signature $(3\big|_{-},8\big|_{+})$. \item 9
Cartan matrices with vanishing determinants, all of which have one
zero eigenvalue and one negative eigenvalue. 7 of these can be
derived from geometric configurations. \item 1 Cartan matrix with
vanishing determinant and with two negative and one zero
eigenvalue. \item 166 Cartan matrices with positive determinants
but with signatures $(2\big|_{-},9\big|_{+})$.
\end{itemize}

\indent Subsequent sections are organized as follows. We begin by
briefly recalling some relevant properties of Coxeter groups and
how they are realized as the Weyl groups of Kac-Moody algebras. In
Section $3$ we display in detail the complete classification of
all rank $10$ Coxeter groups with incidence index $3$ and Coxeter
exponents equal to $2$ or $3$. Section $4$ is devoted to a similar
classification for the rank $11$ case. Finally, we end the paper
with a concluding discussion of our results and directions for
future research.

Discussions of subalgebras of Kac-Moody algebras and in particular
$E_{10(10)}$ along different but related lines may be found in
\cite{Feingold:2003es,Brown:2004,Kleins,BagnoudCarlevaro}.

\section{Coxeter Groups - A Reminder}
The following section gives a very brief overview of some relevant
aspects of Coxeter groups and their geometric realization. More
information may be found in \cite{Humphreys} \cite{Kac}.

A Coxeter group $\mf{C}$ is abstractly defined in terms of
generators $\si_i$ $(i=1,\dots, r)$ that obey the following
relations \beq (\si_i\si_j)^{m_{ij}}=1, \eqnlab{Coxeterrelations}
\eeq where the \emph{Coxeter exponents} $m_{ij}$ are positive
integers with the following properties \beqa m_{ij}&=&m_{ji}
\nn \\
m_{ii}&=&1
\nn \\
m_{ij}&\geq & 2 \quad \mathrm{for} \quad i\neq j.
\eqnlab{Coxeterexponents} \eqa Note that $m_{ii}=1$ implies \beq
\si_i^{2}=1, \eqnlab{reflectionproperty} \eeq which explains why
the Coxeter groups are called \emph{reflection groups}. We focus
in the sequel on the exponents $m_{ij}$ with $i \not=j$. These
Coxeter exponents contain the complete information about the group
$\mf{C}$. It is convenient to encode this information in a
diagram, called a \emph{Coxeter graph}, $\mathbb{G}_{\mf{C}}$. We
associate a node in $\mathbb{G}_{\mf{C}}$ to each reflection
generator $\si_i$. The rank $r$ of $\mf{C}$ is then equal to the
number of nodes in $\cg$. If $m_{ij}=2$, the generators $\si_i$
and $\si_j$ commute, in which case there is no line connecting
nodes $i$ and $j$ in $\cg$. Whenever $m_{ij}>2$ the nodes $i$ and
$j$ are connected by a single line and one writes explicitly the
component $m_{ij}$ over the line, except if $m_{ij}=3$ in which
case the space over the line is left blank.

In order to see the connection with Weyl groups of Kac-Moody
algebras it is useful to exhibit a geometric realization of
$\mf{C}$. This can be achieved by considering the Coxeter group as
a group of linear transformations acting in a vector space, $V$,
of dimension $r=$\hs Rank$\hs \mf{C}$. We will be interested in
\emph{crystallographic} Coxeter groups, which are the ones that
stabilize a lattice, $L$, in $V$. This is the class of Coxeter
groups that appear as Weyl groups of Kac-Moody algebras, and the
lattice $L$ then corresponds to the root lattice of the
Kac-Moody algebra in question. For crystallographic Coxeter
groups, the exponents $m_{ij}$ are restricted to lie in the set
$\{2,3,4,6,\infty\}$.

The action of $\mf{C}$ on $V$ is explicitly given by \beq
\si_i(\al_j)=\al_j-2\f{B(\al_i,\al_j)}{B(\al_i,\al_i)}\al_i,
\eqnlab{geometricreflection} \eeq where $B(\hs,\hs )$ denotes the
scalar product in $V$ and where $\{\al_1,\dots, \al_r\}$ is a basis
of of $V$. The form of \Eqnref{geometricreflection} is precisely
that of a Weyl reflection in the set of simple roots, $\al_i$, of
some Kac-Moody algebra, $\mf{g}$.  The bilinear form $B(\hs, \hs)$
is then the $\mf{g}$-invariant bilinear form restricted to the
Cartan subalgebra. The integers
$$2\f{B(\al_i,\al_j)}{B(\al_i,\al_i)} $$ are called Cartan
integers and form the off-diagonal entries of the Cartan matrix.

In the particular case where the simple roots $\al_i$ of the algebra
$\mf{g}$ all have the same length squared, conveniently taken to be
equal to $2$, the Cartan matrix is symmetric and given by \beq
A_{ij}=B(\al_i,\al_j). \eqnlab{Cartanmatrix} \eeq  The relation
between the Cartan integers and the Coxeter exponents is given (in
the symmetric case) by \cite{Kac} \begin{eqnarray} && A_{ij} = 0 \;
\; \; \; \; \Leftrightarrow \; \; \; m_{ij} = 2, \\ && A_{ij} = -1
\; \; \; \Leftrightarrow \; \; \; m_{ij} = 3 , \\ && A_{ij} < -1 \;
\; \; \Leftrightarrow \; \; \; m_{ij} = \infty .
\end{eqnarray} {}
For the case of connected Kac-Moody algebras with a symmetric
Cartan matrix with only $2$'s, $0$'s and $-1$'s, the Dynkin
diagrams and the Coxeter graphs of the associated Weyl groups
coincide. In order to see this we recall the rules for
constructing the Dynkin diagram associated to a Cartan matrix
$A^{ij}$. For each simple root there corresponds a node in the
diagram. The nodes $i$ and $j$ ($i\neq j$) are connected by
$|A^{ij}|$ lines. Thus for the cases under consideration we have
always $|A^{ij}|=1$ or $|A^{ij}|=0$ so the Dynkin diagram indeed
coincides with the Coxeter graph. Hence, in this case they carry
the same information and instead of classifying all Coxeter
exponents $m_{ij}$ we perform the classification of all Cartan
matrices whose Dynkin diagrams have the desired structure. The
result is the same.

Note that even though the Cartan matrix of $\mf{g}=E_{10(10)}$ is
symmetric and has only $0$ and $-1$ off the diagonal, the Cartan
matrix of regular subalgebras of $E_{10(10)}$ constructed along the
lines of \cite{Feingold:2003es}, while necessarily symmetric, might
have integers $<-1$ off the diagonal. The corresponding Coxeter
exponents might then be equal to infinity. This is because the
scalar product of real, (non simple) positive roots of $E_{10(10)}$
might be $<-1$.  For instance, even if one restricts one's attention
to symmetry, electric, magnetic or gravitational roots, one finds that the
scalar products are in the set $\{2, 1, 0, -1,-2,-3\}$ with $-3$
reached for the scalar products of some gravitational roots among
themselves. Let us show this explicitly for a few examples.

Consider the roots, $\al$, in the $\beta$-space basis
\cite{Damour:2002et}, where they are written as linear forms on
the Cartan subalgebra, i.e. \beq
\al(\be)=\sum_{i=1}^{10}\al_i\be^{i}, \eeq where a general element
of the Cartan subalgebra is $h=\be^{i}\al^{\vee}_i$. The metric in
the root space of $E_{10(10)}$ is Lorentzian and takes the
following form \beq (\al |\al^{\prime})=\sum_{i=1}^{10}\al_i
\al^{\prime}_i-\frac{1}{9}\Big(\sum_{i=1}^{10}\al_i\Big)\Big(\sum_{j=1}^{10}\al^{\prime}_j\Big).
\eeq To illustrate the procedure, we choose arbitrarily two
magnetic roots (level 2) of the form \beq
\al_1^{M}(\be)=\be^1+\be^2+\be^3+\be^4+\be^5+\be^6; \qquad
\al_2^{M}(\be)=\be^1+\be^2+\be^7+\be^8+\be^9+\be^{10}, \eeq and
using the bilinear form $(\ |\ )$ one may check that the scalar
product between them is \beq (\al^{M}_1|\al^{M}_2)=-2. \eeq By
ascending to gravitational roots at level 3 one finds scalar
products with lowest negative value being $-3$. For example, the
two roots \beq \al_{1}^{G}(\be)=2\be^1+\be^2+\dots +\be^8 \quad
\mathrm{and} \quad \al_{2}^{G}(\be)=\be^2+\dots
+\be^7+2\be^9+\be^{10}, \eeq are real and their scalar product is
\beq (\al_1^{G}|\al_2^{G})=-3. \eeq The same procedure can be
applied for real roots at any level yielding scalar products
taking lower and lower negative integer numbers.

In principle one can obtain arbitrarily negative scalar products
by choosing real roots of higher and higher level. Coxeter
subgroups of the Weyl group of $E_{10(10)}$ with Coxeter exponents
equal to $2$ or $3$ (and not $\infty$) are thus rather special.

\section{Classification of Rank $10$ Coxeter Groups with $\mc{I}=3$}

Let $\cg$ be a rank $10$ Coxeter graph with $\mc{I}=3$ and
$m_{ij}= 2,3$ ($i \not=j$). The structure of $\cg$ is completely
encoded in its associated Cartan matrix, $A_{\cg}$, which is a
$10\times 10$ symmetric matrix with $2$ on each diagonal entry and
zeroes on the off-diagonal except for exactly three entries in
each row (and column) which are equal to $-1$. If an off-diagonal
entry is non-vanishing, the corresponding nodes in $\cg$ are
connected by a single line. Similarly, a vanishing off-diagonal
entry in $A_{\cg}$ implies that the associated nodes in $\cg$ are
disconnected. Our task is now to find all inequivalent such
matrices.  We shall not give the details of the computations here
but shall only outline the derivation.

We construct the matrices row by row as follows. A choice has to
be made for the first row but any choice fulfilling the
aforementioned requirements is acceptable. We denote a general
Cartan matrix by $C$ and take without generality the first row to
be \beq [C_{1i}]=\left(
\begin{array}{cccccccccc}
\ph{-}2 &-1 & -1 & -1 & \ph{-}0 & \ph{-}0 & \ph{-}0 & \ph{-}0 & \ph{-}0 & \ph{-}0 \\
\end{array}\right),
\eqnlab{firstrow}
\eeq
where $i=1,\dots, 10$. Because $C$ is symmetric the next row must start as
\beq
[C_{2i}]=\left( \begin{array}{ccc}
-1 & \ph{-}2 & \dots  \\
\end{array}\right),
\eqnlab{secondrow} \eeq and we know that two more entries must be
equal to $-1$. There are only three distinct choices. To
understand this it is helpful to keep the Coxeter graph in mind.
The first row tells us that node $1$ is connected to nodes $2,3$
and $4$. As we take the next step we want to determine the number
of distinct connections to node $2$. The possibilities are: (i)
node $2$ is further connected to $3$ and $4$, (ii) node $2$ is
connected to $3$ or $4$ and then to any of the nodes in the set
$\{5,\dots, 10\}$, (iii) node $2$ is connected to two of the nodes
in the set $\{5,\dots, 10\}$. Hence, we arrive at \beqa
{}[C_{2i}]_{(\textbf{1})}&=& \left( \begin{array}{cccccccccc}
-1 & \ph{-}2 & -1 & -1 & \ph{-}0 & \ph{-}0 & \ph{-}0 & \ph{-}0 & \ph{-}0 & \ph{-}0 \\
\end{array}\right)
\\
{}[C_{2i}]_{(\textbf{2})}&=&\left( \begin{array}{cccccccccc}
-1 & \ph{-}2 & -1 & \ph{-}0 & -1 & \ph{-}0 & \ph{-}0 & \ph{-}0 & \ph{-}0 & \ph{-}0 \\
\end{array}\right)
\\
{}[C_{2i}]_{(\textbf{3})}&=&\left( \begin{array}{cccccccccc}
-1 & \ph{-}2 & \ph{-}0 & \ph{-}0 & -1 & -1 & \ph{-}0 & \ph{-}0 & \ph{-}0 & \ph{-}0 \\
\end{array}\right).
\eqnlab{secondrow123} \eqa We now proceed to the third row. For
$C_{(\tbf{1})}$ and $C_{(\tbf{2})}$, the off-diagonal components
$\{3,1\}$ and $\{3,2\}$ are already fixed to $-1$ by symmetry. We
must find the inequivalent ways to add the third non-vanishing
entry. $C_{(\tbf{1})}$ admits two possibilities:
$[C_{34}]_{(\tbf{1})}=-1$ or $[C_{35}]_{(\tbf{1})}=-1$. For
$C_{(\tbf{2})}$, there are three distinct
choices for the entry $-1$, namely  $[C_{34}]_{(\tbf{2})}, [C_{35}]_{(\tbf{2})}$ or $[C_{36}]_{(\tbf{2})}$.\\
\indent In the third case, $C_{(\tbf{3})}$, only the first entry of
the third row is determined by symmetry. This gives us five different
possibilities for the distribution of two non-vanishing off-diagonal entries:
\beqa
 {}[C_{3i}]_{(\textbf{3})_{1}}&=&\left( \begin{array}{cccccccccc}
-1 & \ph{-}0 & \ph{-}2 & -1 & -1 & \ph{-}0 & \ph{-}0 & \ph{-}0 & \ph{-}0 & \ph{-}0 \\
\end{array}\right)
\\
 {}[C_{3i}]_{(\textbf{3})_{2}}&=&\left( \begin{array}{cccccccccc}
-1 & \ph{-}0 & \ph{-}2 & -1 & \ph{-}0 & \ph{-}0 & -1 & \ph{-}0 & \ph{-}0 & \ph{-}0 \\
\end{array}\right)
\\
 {}[C_{3i}]_{(\textbf{3})_{3}}&=&\left( \begin{array}{cccccccccc}
-1 & \ph{-}0 & \ph{-}2 & \ph{-}0 & -1 & -1 & \ph{-}0 & \ph{-}0 & \ph{-}0 & \ph{-}0 \\
\end{array}\right)
\\
 {}[C_{3i}]_{(\textbf{3})_{4}}&=&\left( \begin{array}{cccccccccc}
-1 & \ph{-}0 & \ph{-}2 & \ph{-}0 & -1 & \ph{-}0 & -1 & \ph{-}0 & \ph{-}0 & \ph{-}0 \\
\end{array}\right)
\\
 {}[C_{3i}]_{(\textbf{3})_{5}}&=&\left( \begin{array}{cccccccccc}
-1 & \ph{-}0 & \ph{-}2 & \ph{-}0 & \ph{-}0 & \ph{-}0 & -1 & -1 & \ph{-}0 & \ph{-}0 \\
\end{array}\right).
\eqnlab{thirdrow3} \eqa At this point we thus have ten different
cases to consider. Repeating the same procedure we find $33$
possibilities after adding the fourth row, $98$ possibilities
after the fifth row, $296$ after the sixth and $574$ possibilities
after the seventh row.

The remaining task is to add rows eight, nine and ten. These last
steps will actually considerably restrict the number of distinct
possibilities because many of the matrices we have found so far
cannot be extended up to rank $10$ in such a way as to preserve
the condition $\mc{I}=3$: the construction might get obstructed.
First of all, we know that the tenth row will be completely fixed
by symmetry. Hence, we must find the various possibilities for
rows eight and nine. Only the entries $[C_{89}], [C_{8(10)}]$ and
$[C_{9(10)}]$ are undetermined by symmetry. To this end we
consider the triple $(s_8,s_9,s_{10})$, where $s_8,s_9$ and
$s_{10}$ denote the three sums \beq s_k=\sum_{i=1}^{7}\vert
[C_{ki}]\vert  \qquad k=8,9,10. \eqnlab{sums} \eeq The various
possibilities for the values of the triple, $(s_8,s_9,s_{10})$,
fixes the distinct choice for the remaining entries $[C_{89}],
[C_{8(10)}]$ and $[C_{9(10)}]$ according to \beq
\begin{array} {rcccl}
&s_{8} &s_{9} &s_{10}&\\
&1&1&1&\\
&1&2&2&\\
&2&1&2&\\
&2&2&1&\\
&2&2&3&\\
&2&3&2&\\
&3&2&2&\\
&3&3&3&
\end{array}
\qquad
\begin{array} {rcccl}
&[C_{98}] & [C_{8(10)}] & [C_{9(10)}] &\\
&-1&-1&-1&\\
&-1&-1& \ph{-}0&\\
&-1& \ph{-}0&-1&\\
& \ph{-}0&-1&-1&\\
&-1& \ph{-}0& \ph{-}0&\\
& \ph{-}0&-1& \ph{-}0&\\
&\ph{-}0&\ph{-}0&-1&\\
&\ph{-}0&\ph{-}0&\ph{-}0.&
\end{array}
\eqnlab{triples} \eeq All other choices of $(s_{8},s_9,s_{10})$ are
incompatible with having incidence index $3$. {}For instance the
first two $1$'s in the triple $(1,1,2)$ forces $[C_{89}] =
[C_{8(10)}]= [C_{9(10)}] = -1$ with the resulting contradiction that
there are four $-1$'s on the last line. Therefore, out of the $574$
Cartan matrices we had up to seven rows, we find that only $256$ of
them allow for an extension to rank $10$ in accordance with
\Eqnref{triples}. Among these $10\times 10$ matrices, $109$ are
Lorentzian (i.e. have negative determinant), 12 are degenerate (one
null, one negative and eight positive eigenvalues) and $135$ of them
have positive determinant but come with signature
$(2\big|_-,8\big|_{+})$.

The final step consists of determining how many of these matrices
are actually equivalent up to a permutation of the vertices in the
associated Coxeter graphs, or, in other words, up to a simultaneous
exchange of rows and columns so as to preserve the occurrence of $2$
on the diagonal. We have done this explicitly and found that many of
the matrices obtained so far were in fact equivalent.  We obtained
that in addition to the $6$ Lorentzian cases discovered in
\cite{GeometricConf} there are only $3$ distinct Cartan matrices
with Lorentzian signature. Furthermore, in \cite{GeometricConf} we
uncovered one case which was degenerate and it turns out that only
one additional case exists. Finally, out of the $135$ matrices with
positive determinants, $10$ are distinct (and have signature with 8
$+$'s and 2 $-$'s). The result is thus that there exist $21$ rank
$10$ Coxeter groups, or, equivalently, $21$ rank $10$ indefinite
Kac-Moody algebras, with incidence index $3$. The corresponding
Coxeter graphs are reproduced in Tables $1-4$.

\section{Classification of Rank $11$ Coxeter Groups with $\mc{I}=4$}

The particular class of rank $10$ Coxeter groups discussed above
were discovered by examining geometric configurations of the type
$(n_m,g_3)=(10_3,10_3)$. A similar analysis can be done for the
geometric configurations with $11$ points and $11$ lines,
$(n_m,g_3)=(11_3,11_3)$. There exist $31$ such configurations
\cite{Page} and each of these gives rise, through its
line-incidence diagram, to a Coxeter subgroup of the Weyl group of
$E_{11(11)}$, whose significance in the context of M-theory was
first pointed out in \cite{West}. The procedure is identical to
the one performed for the $(10_3,10_3)$-configurations in
\cite{GeometricConf}. This is easy to understand given the fact
that the field contents for $E_{10(10)}$ and $E_{11(11)}$ at low
levels are the same. All $(11_3,11_3)$-configurations have the
property that each line is parallel to four other lines and hence
the associated Coxeter graphs have incidence index $\mc{I}=4$.
Moreover the Coxeter exponents lie in the set $\{2,3\}$.

Following the procedure outlined in Section $3$ we have classified
all rank $11$ Coxeter groups with these properties. The end result
is that there exist $252$ rank $11$ Coxeter groups with
$\mc{I}=4$. Of  these, 28 can be obtained from geometric
configurations and so correspond to subgroups of the Weyl group of
$E_{11(11)}$.

Some new features arise in comparison to the rank $10$ case: (i)
also Cartan matrices with non-vanishing determinants come with a
degeneracy, (ii) one of the Cartan matrices with vanishing
determinant have two negative eigenvalues, (iii) some of the
Cartan matrices with negative determinants have signatures
$(3\big|_{-},8\big|_{+})$. The various classes of Cartan matrices
were listed in Section 1 so we will not repeat it here.

Because of obvious restrictions of space we do not exhibit the
full list of rank $11$ Coxeter graphs. However, in Table $5$ we
display some selected graphs that do not arise as line-incidence
diagrams of geometric configurations. All the 252 Cartan matrices
are assembled in the file ``Coxeter11-4.nb'' which is included in
``Coxeter.zip'' that can be downloaded from the database arXiv.org
of Cornell University \cite{link}.\footnote{Since the first version of this article, the file ``Coxeter11-4.nb'' has been replaced by ``Coxeter11-4v4.nb'' that contains the correct number, 266, of inequivalent rank 11 Cartan matrices with incidence index 4. See Errata on page 19.}

\section{Conclusions}
In this note we have extended the results of \cite{GeometricConf}
by classifying all rank $10$ Coxeter groups with incidence index
$\mc{I}=3$ and Coxeter exponents $m_{ij}=2,3$ ($i \not=j$),
including those that are not determined by geometric
configurations. We find that except for the $7$ cases that were
uncovered in \cite{GeometricConf}, there exist $14$ additional
Coxeter groups with similar properties. Among those, 10 cannot be
regularly embedded in $E_{10(10)}$ since they do not have the correct
signature. Although we do not know if the other Coxeter groups
correspond to subgroups of the Weyl group of $E_{10(10)}$, one might
speculate that they can perhaps be associated with some other
class of subalgebras of $E_{10(10)}$ that goes beyond the ``electric''
subalgebras previously considered.

The natural starting point for such an investigation would be to
ascend to level $2$ in the decomposition of $E_{10(10)}$ and
investigate ``magnetic'' subalgebras. These correspond to
geometric configurations of the type $(n_m,g_6)$, i.e. with $6$
points on each line. It is possible that some of the rank $10$
Coxeter groups derived in this way could have $\mc{I}=3$ and
Coxeter exponents $m_{ij}=2,3$ and would thus fall into the
classification scheme of this paper.

We have also classified all rank $11$ Coxeter groups with
incidence index $\mc{I}=4$. The analysis revealed $252$ such
Coxeter groups, including 28 that can be obtained from
$(11_3,11_3)$ configurations and so can be regularly embedded in
$E_{11(11)}$.

We should mention here the following related fact.  Because the
exponent $\infty$ does not occur among the Coxeter exponents, any
pair of elements in the set of generating reflections generates a
finite group (the group is called ``2-spherical" \cite{Caprace}).
2-spherical Coxeter subgroups of $E_{10(10)}$ and $E_{11(11)}$ are
rather rare - in fact, up to conjugation, there are only a finite
number of them \cite{Caprace}.  It would be of interest to
determine them all.

\vspace{.3cm}

\noindent \textbf{Acknowledgments:} MH is grateful to the organizers
of the ``Fourth International School on Field Theory and
Gravitation'', held in Rio de Janeiro in May 2006, for their kind
hospitality and for the opportunity to lecture on cosmological
billiards. We thank Pierre-Emmanuel Caprace, Arjan Keurentjes and
Bernhard M\"uhlherr for discussions at various stages in the
preparation of this article.  Work supported in part by IISN-Belgium
(convention 4.4511.06 (M.H. and P.S) and convention 4.4505.86 (M.H.
and D.P)), by the Belgian National Lottery, by the European
Commission FP6 RTN programme MRTN-CT-2004-005104 (M.H., M.L. and
D.P.), and by the Belgian Federal Science Policy Office through the
Interuniversity Attraction Pole P5/27.  Mauricio Leston was also
supported in part by the ``FWO-Vlaanderen'' through project
G.0428.06 and by a CONICET graduate scholarship.

\begin{center}
\begin{table}
  \begin{tabular}{ |m{65mm}|m{10mm}|m{65mm}|m{10mm}|}
\hline
 Coxeter Graph $\mbb{C}_{\mf{C}}$ & $\det \hs C$ & Coxeter Graph  $\mbb{C}_{\mf{C}}$ & $\det\hs C$ \\
    \hline
     \hline
     \includegraphics[width=60mm]{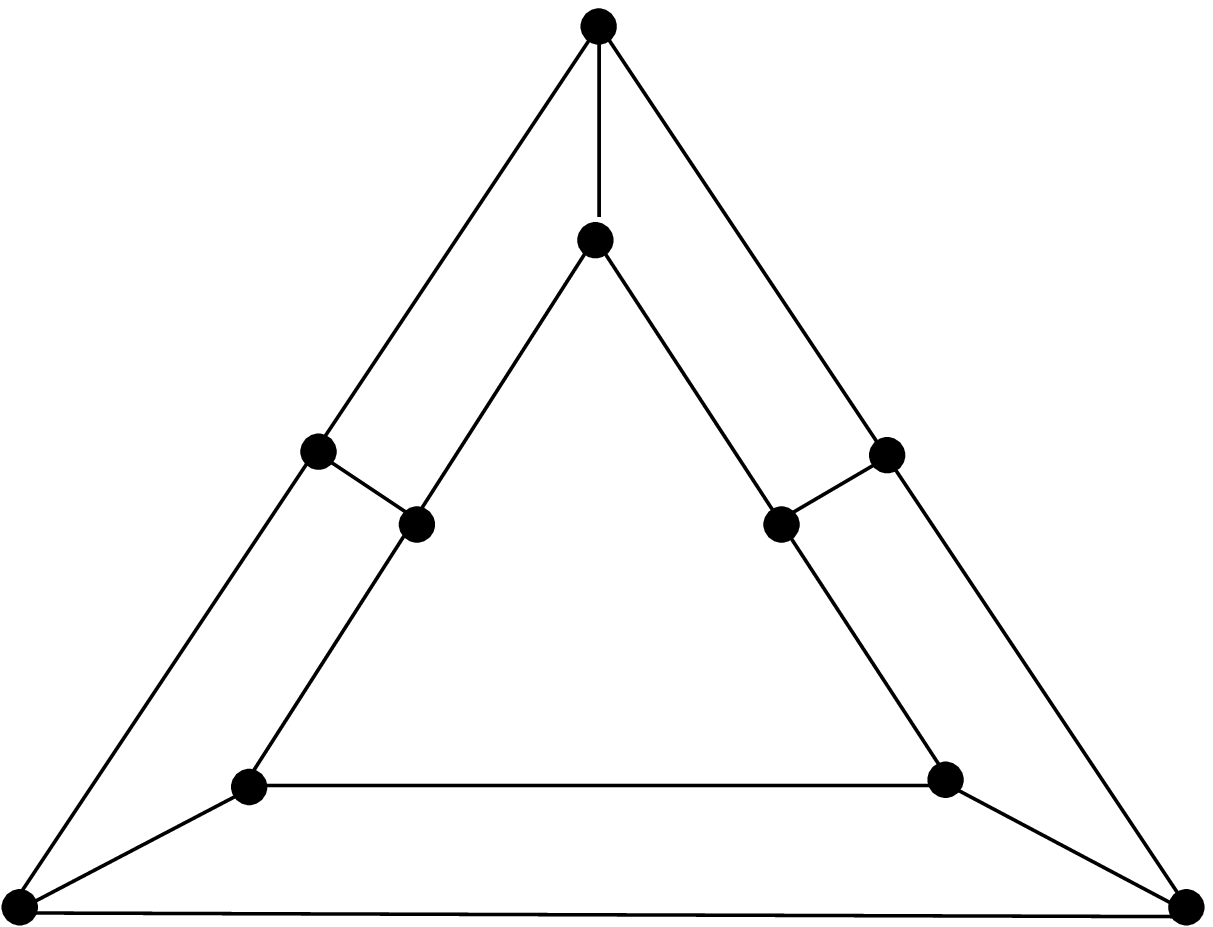} & $-121$ &
      \includegraphics[width=60mm]{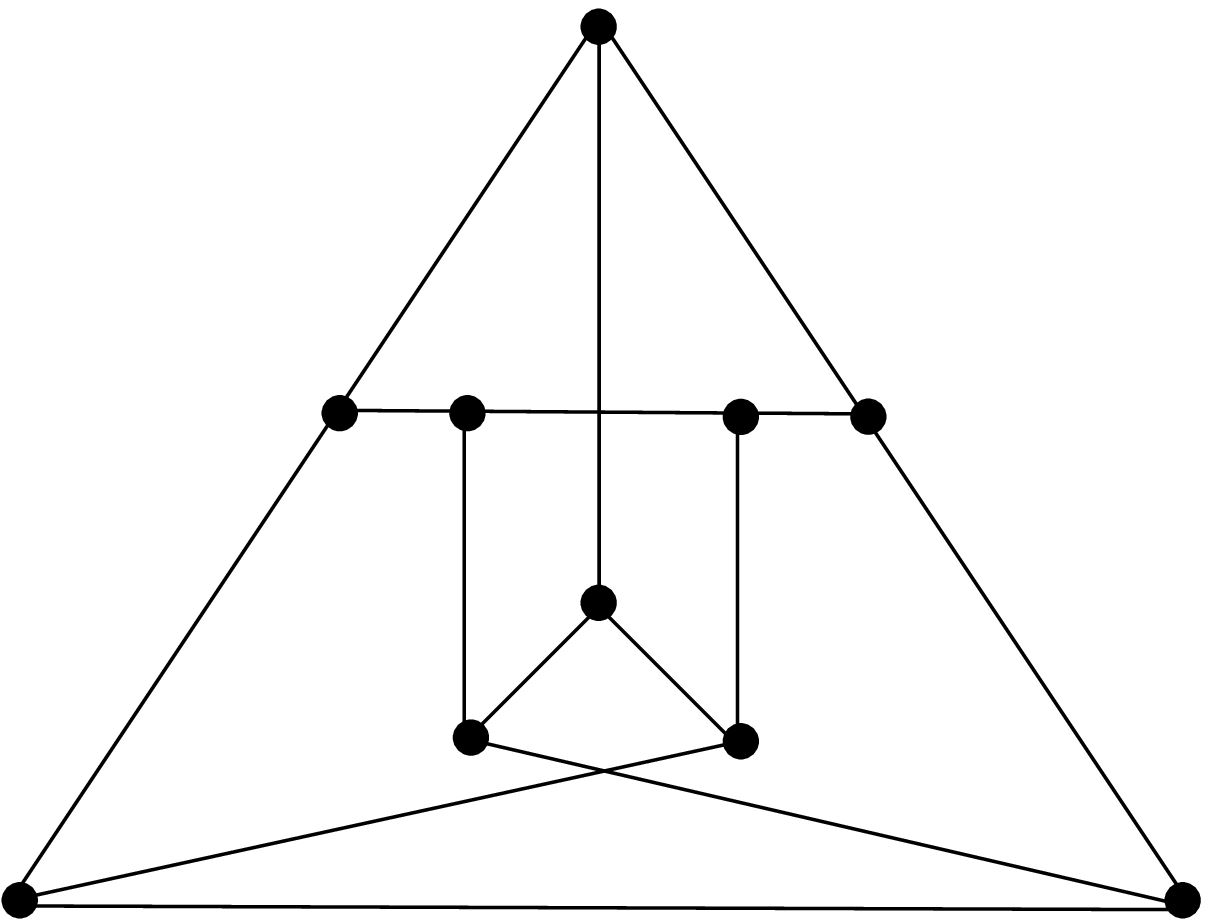} & $ -256 $  \\
    \hline
      \includegraphics[width=60mm]{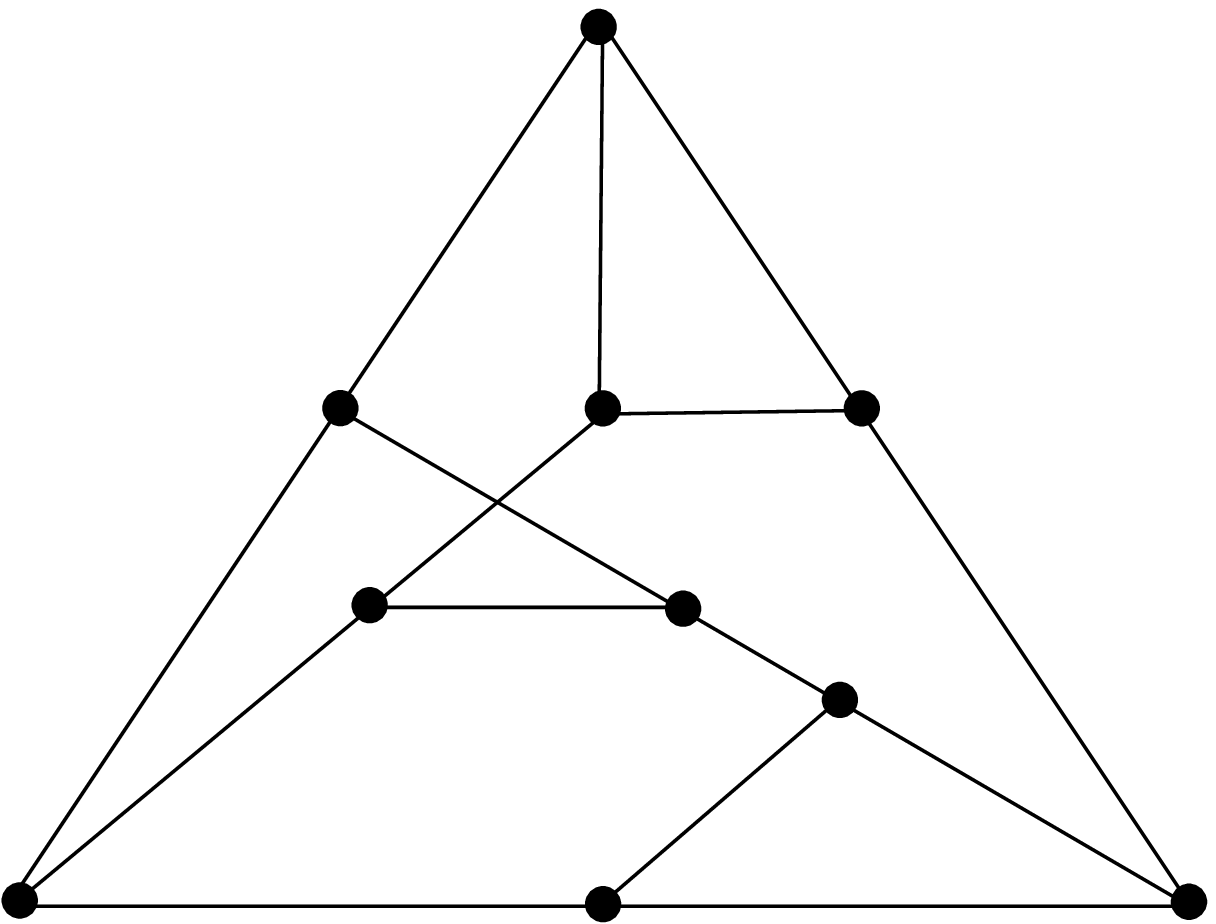} & $-25$ &
      \includegraphics[width=60mm]{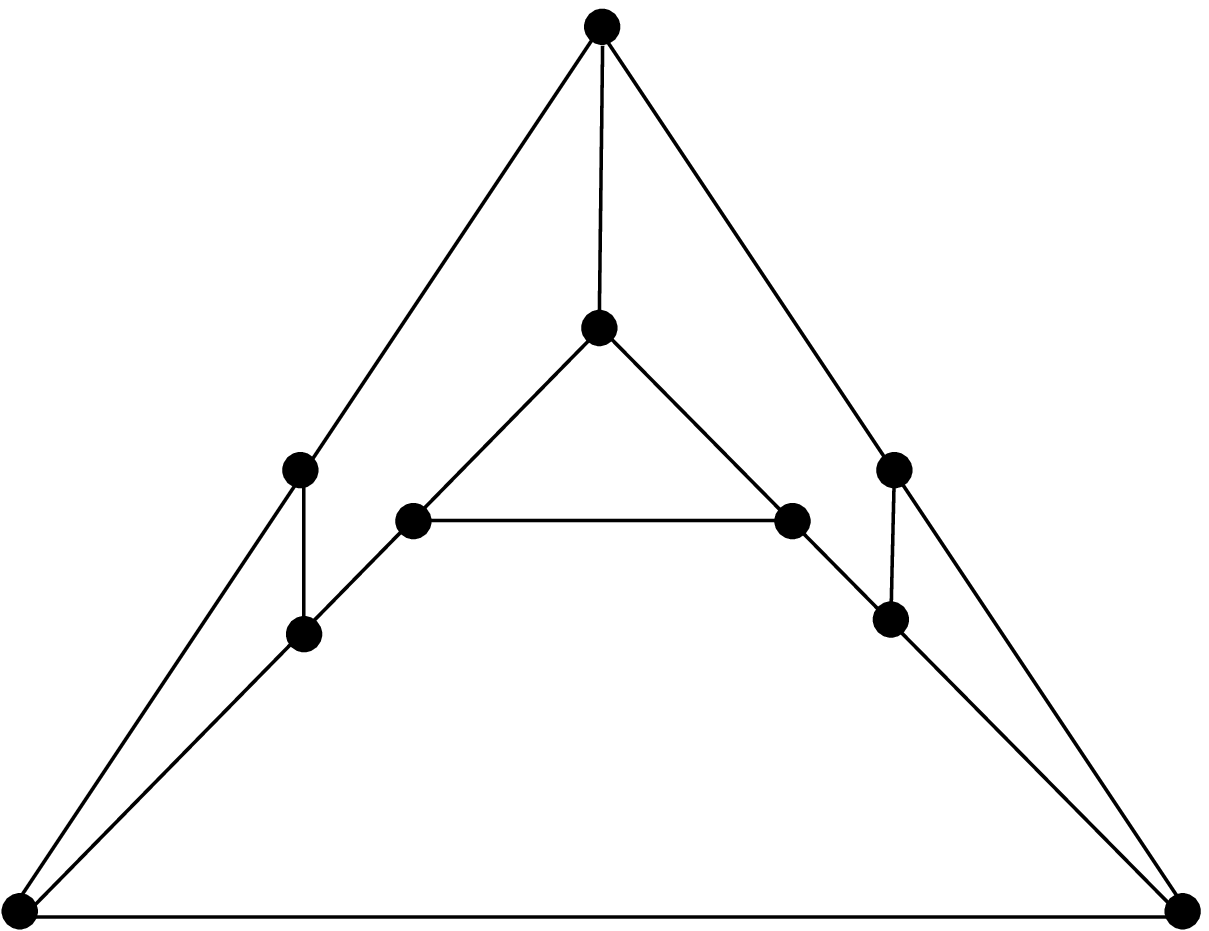} & $-16$ \\
      \hline
       \includegraphics[width=60mm]{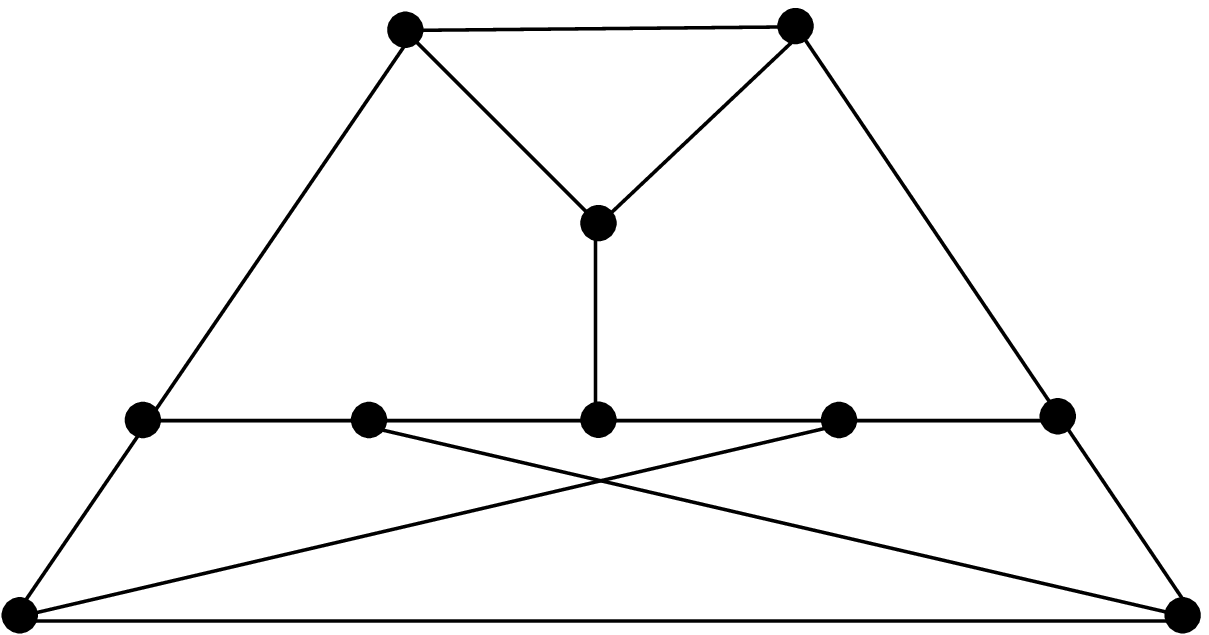} & $-64$ &
      \includegraphics[width=60mm]{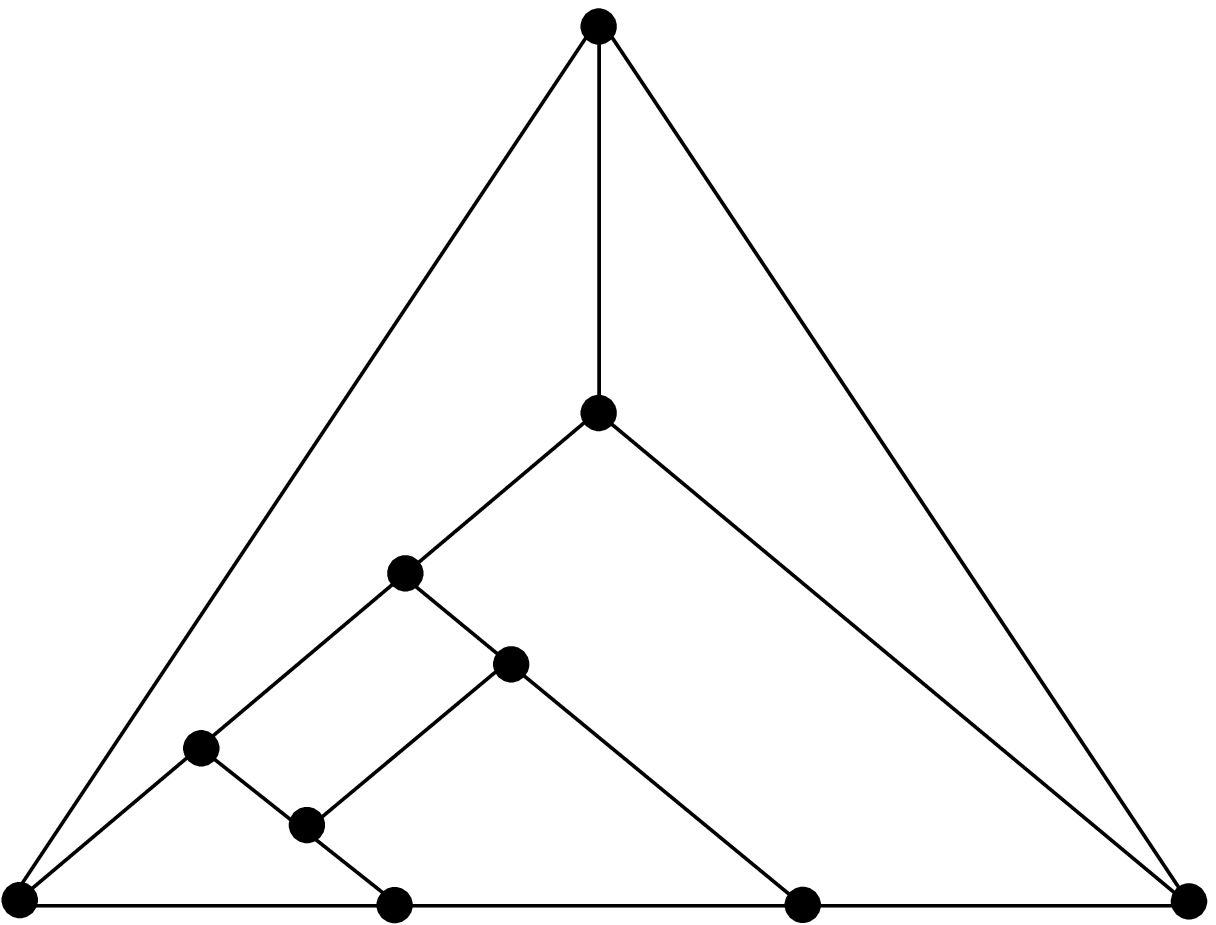} & $-49$ \\
          \hline
      \end{tabular}
         \caption{Lorentzian rank $10$ Coxeter graphs with incidence index $\mc{I}=3$ and Coxeter exponents $2$ or $3$. All
         of these graphs correspond to line-incidence diagrams of geometric configurations of the type $(n_m,g_3)$.}
\end{table}
%\label{table:finite}
\end{center}

\begin{center}
\begin{table}
  \begin{tabular}{ |m{65mm}|m{10mm}|m{65mm}|m{10mm}|}
\hline
 Coxeter Graph $\mbb{C}_{\mf{C}}$ & $\det \hs C$ & Coxeter Graph  $\mbb{C}_{\mf{C}}$ & $\det\hs C$ \\
    \hline
 \hline
      \includegraphics[width=60mm]{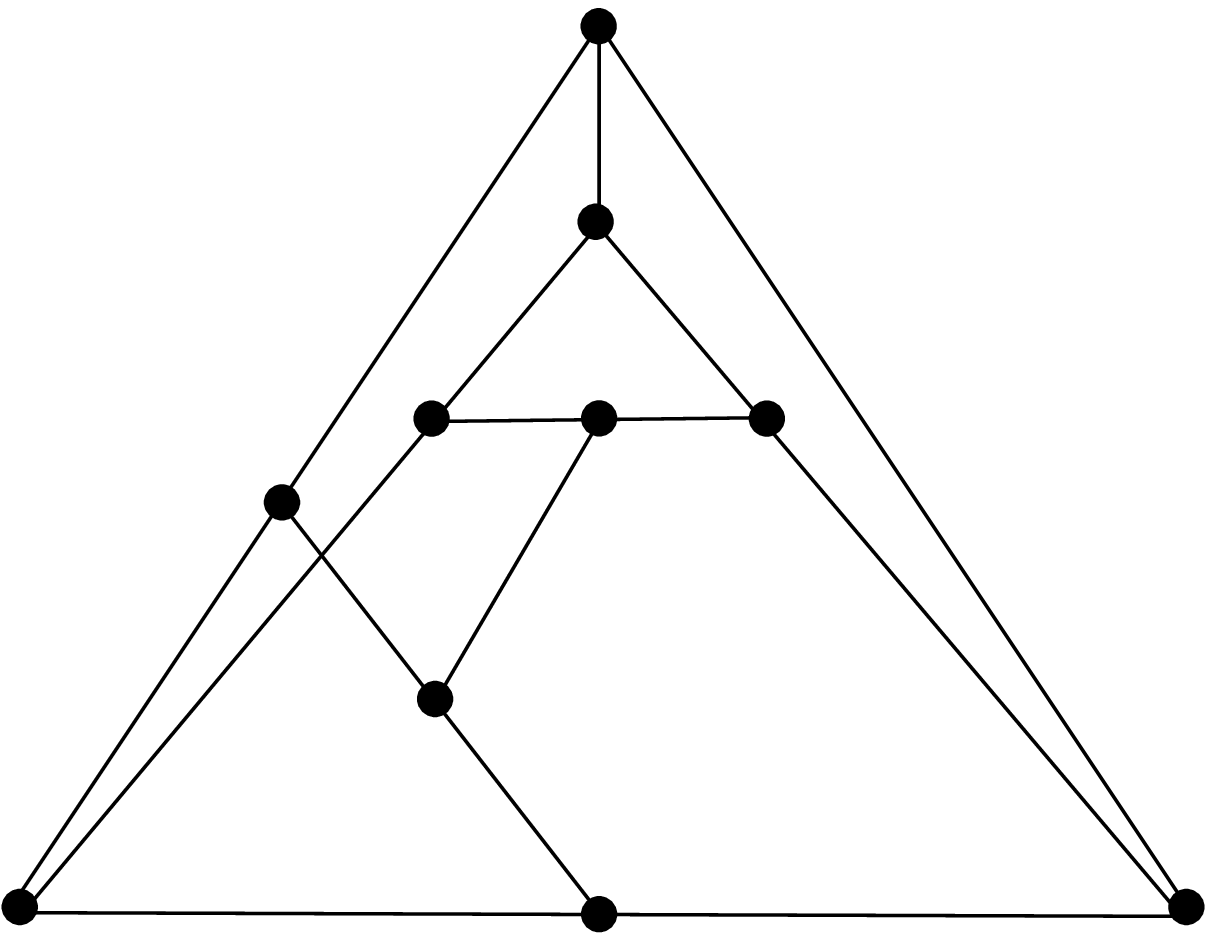} & $-165$ &
      \includegraphics[width=60mm]{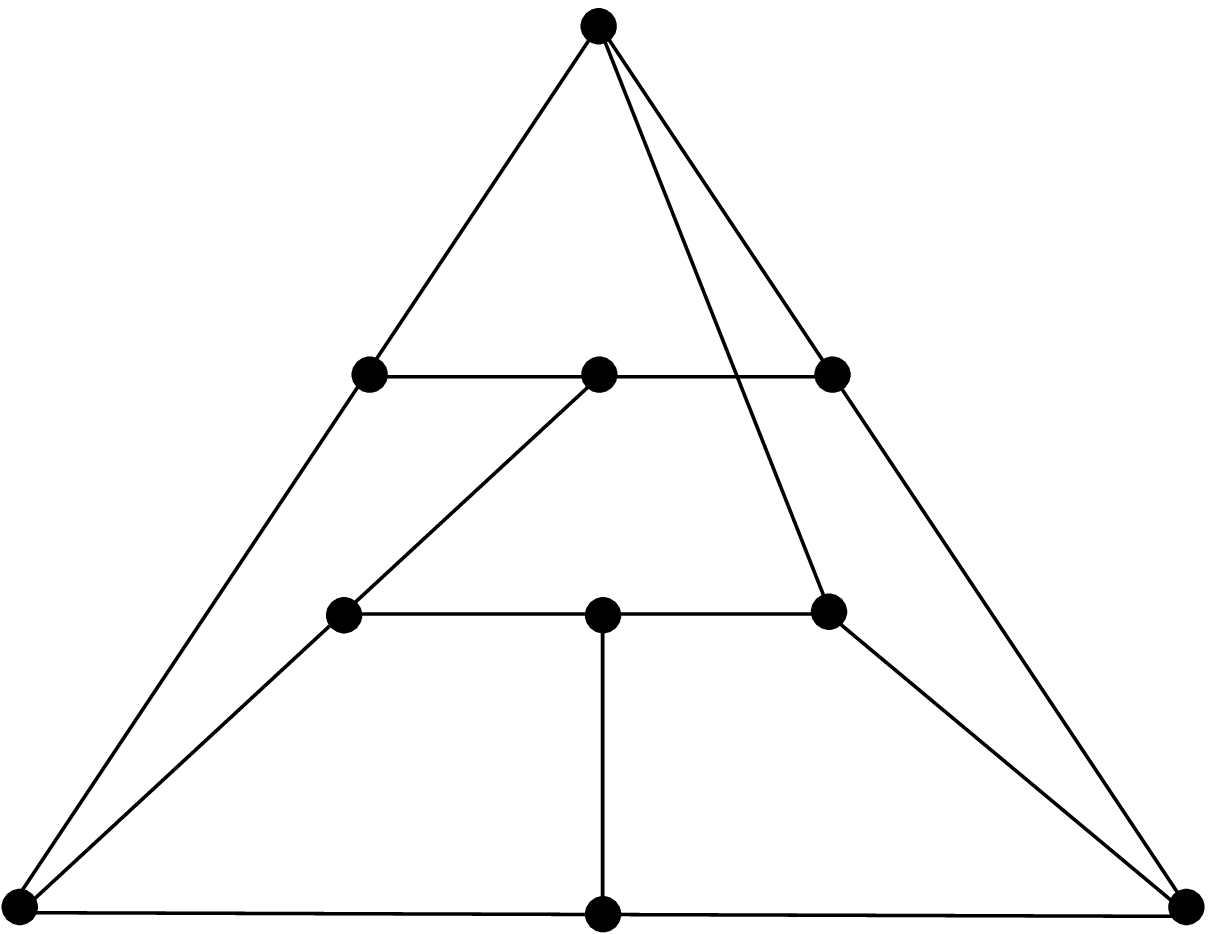} & $-125$\\
      \hline
     \includegraphics[width=60mm]{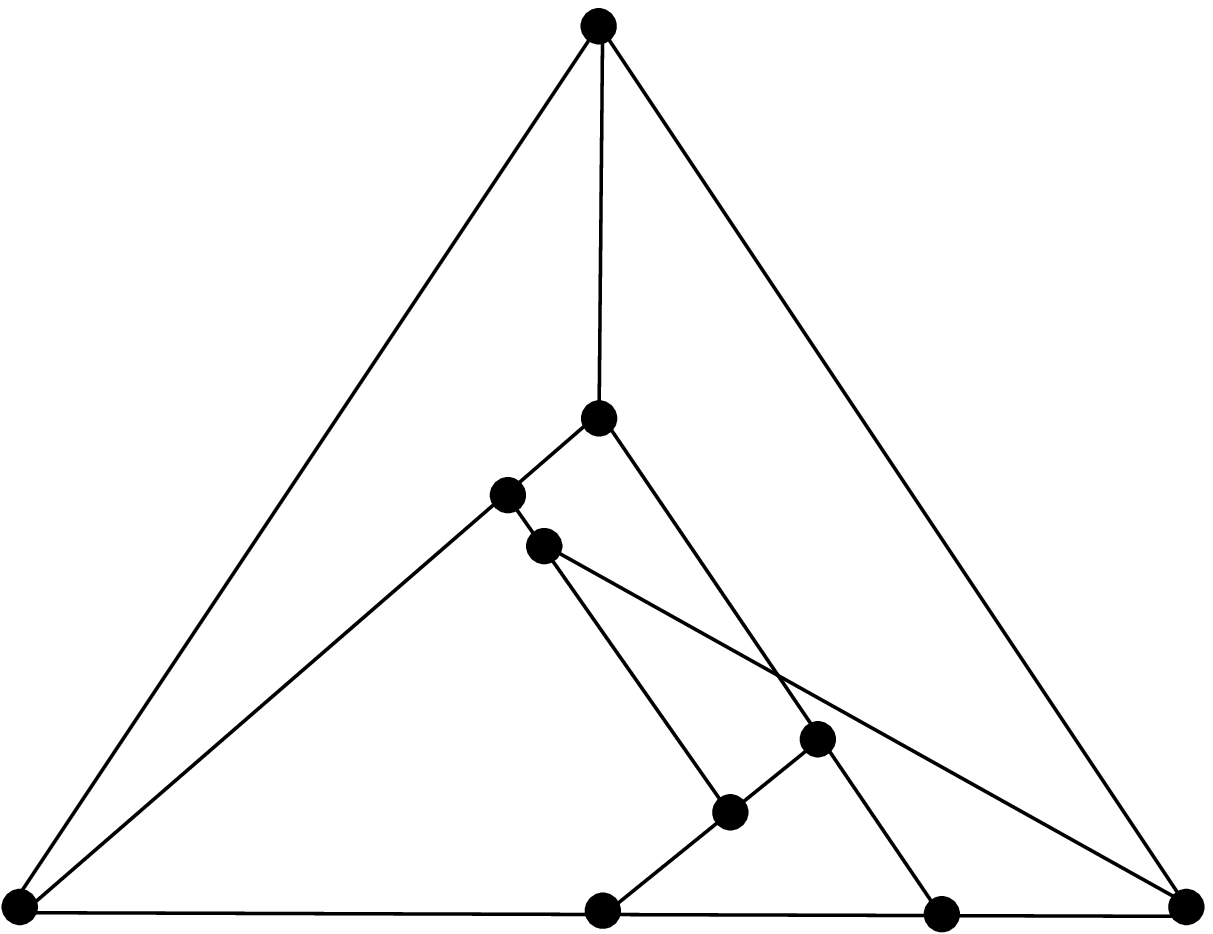} & $-192$
     &  {}  & {}\\
      \hline
  \end{tabular}
         \caption{Lorentzian rank $10$ Coxeter graphs with incidence index $\mc{I}=3$ and Coxeter exponents $2$ or $3$.
         None of these graphs correspond to line-incidence diagrams of geometric configurations of the type $(n_m,g_3)$.}
\end{table}
%\label{table:finite}
\end{center}

\begin{center}
\begin{table}
  \begin{tabular}{ |m{65mm}|m{10mm}|m{65mm}|m{10mm}|}
\hline
 Coxeter Graph $\mbb{C}_{\mf{C}}$ & $\det \hs C$ & Coxeter Graph  $\mbb{C}_{\mf{C}}$ & $\det\hs C$ \\
    \hline
 \hline
      \includegraphics[width=60mm]{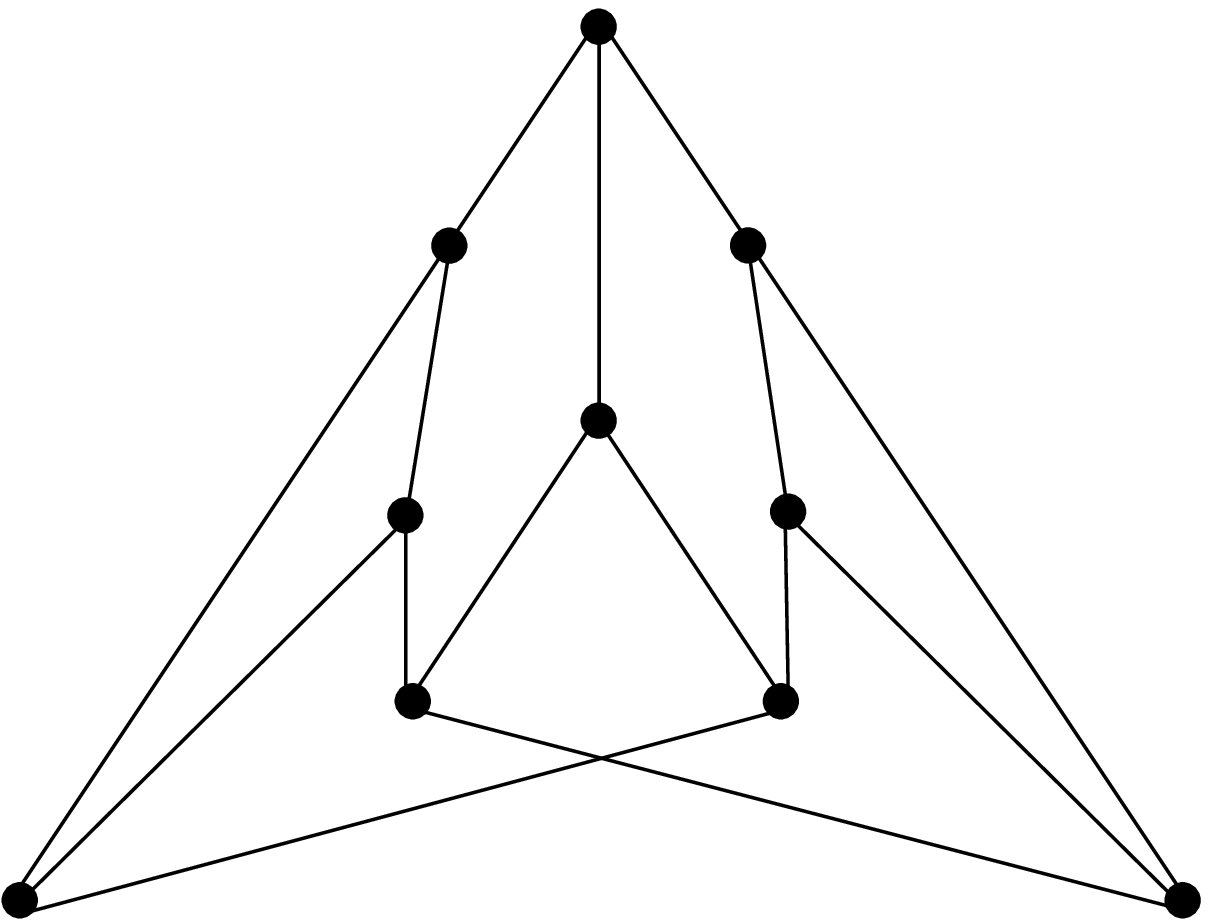} & $ 0$ &
      \includegraphics[width=60mm]{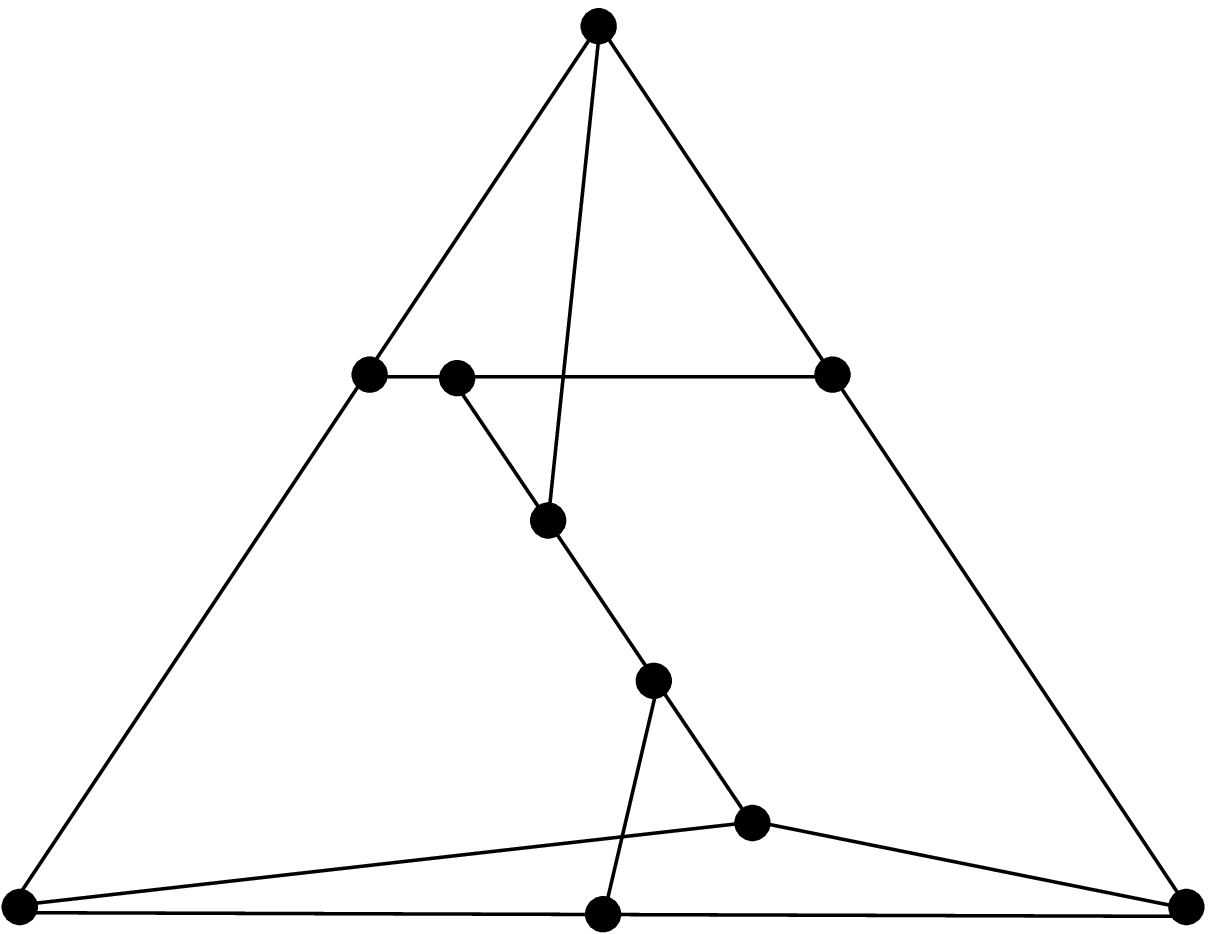} & $0$\\
      \hline
          \hline
  \end{tabular}
         \caption{Degenerate rank $10$ Coxeter graphs with incidence index $\mc{I}=3$ and Coxeter exponents $2$ or $3$.
         Only the leftmost graph correspond the line-incidence diagram of a
         geometric configuration of the type $(n_m,g_3)$.}
\end{table}
%\label{table:finite}
\end{center}

\begin{center}
\begin{table}
  \begin{tabular}{ |m{65mm}|m{10mm}|m{65mm}|m{10mm}|}
\hline
 Coxeter Graph $\mbb{C}_{\mf{C}}$ & $\det \hs C$ & Coxeter Graph  $\mbb{C}_{\mf{C}}$ & $\det\hs C$ \\
    \hline
     \hline
     \includegraphics[width=60mm]{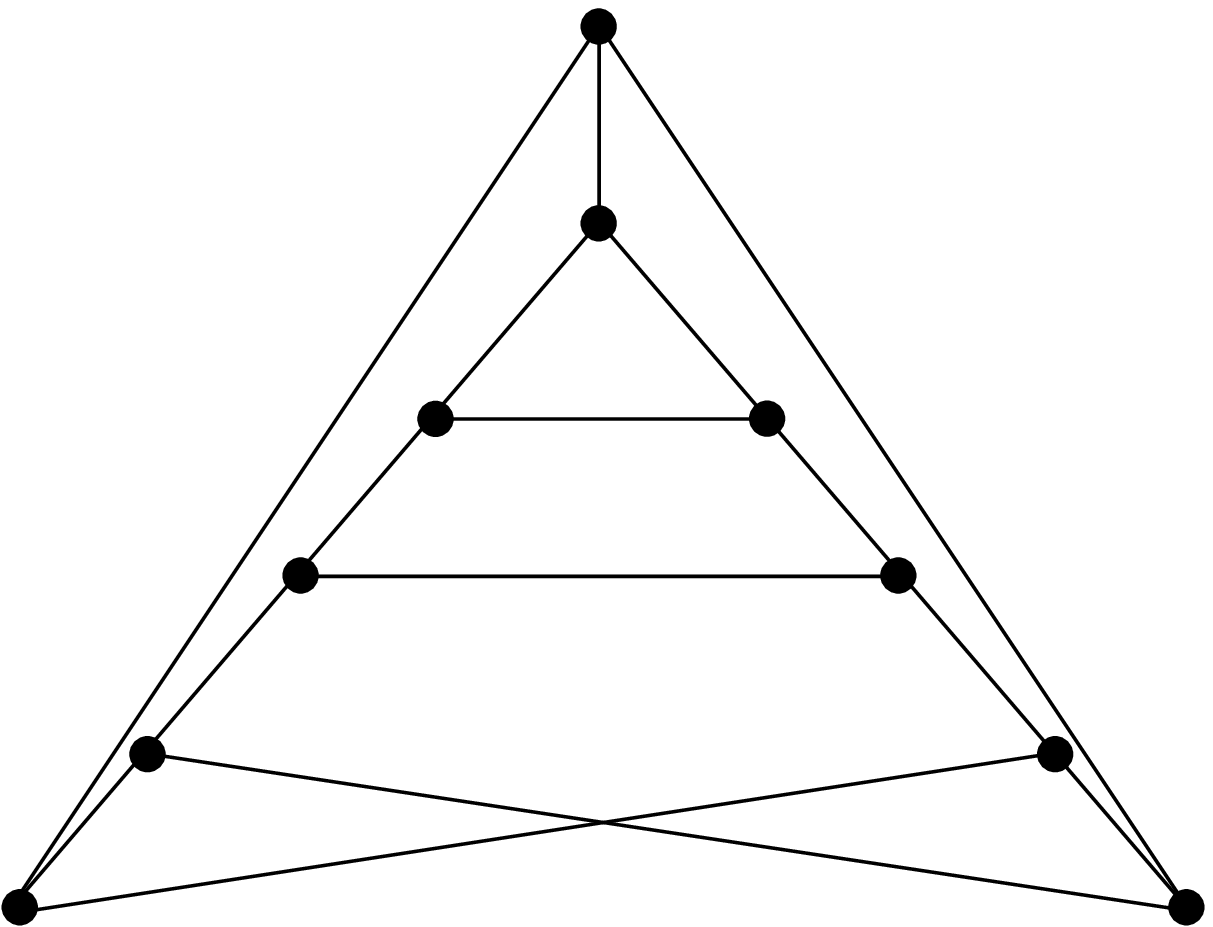} & $52$ &
      \includegraphics[width=60mm]{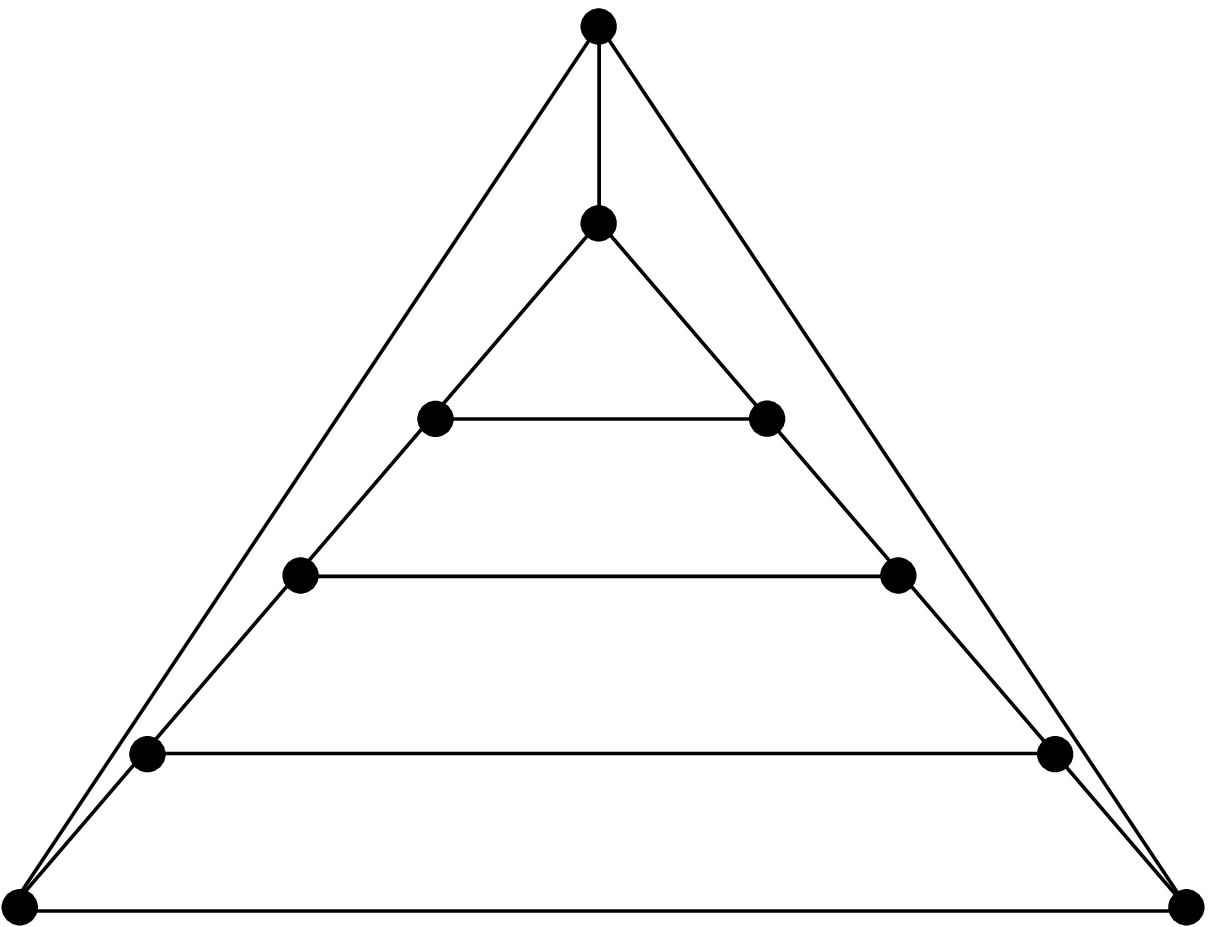} & $ 55 $  \\
    \hline
      \includegraphics[width=60mm]{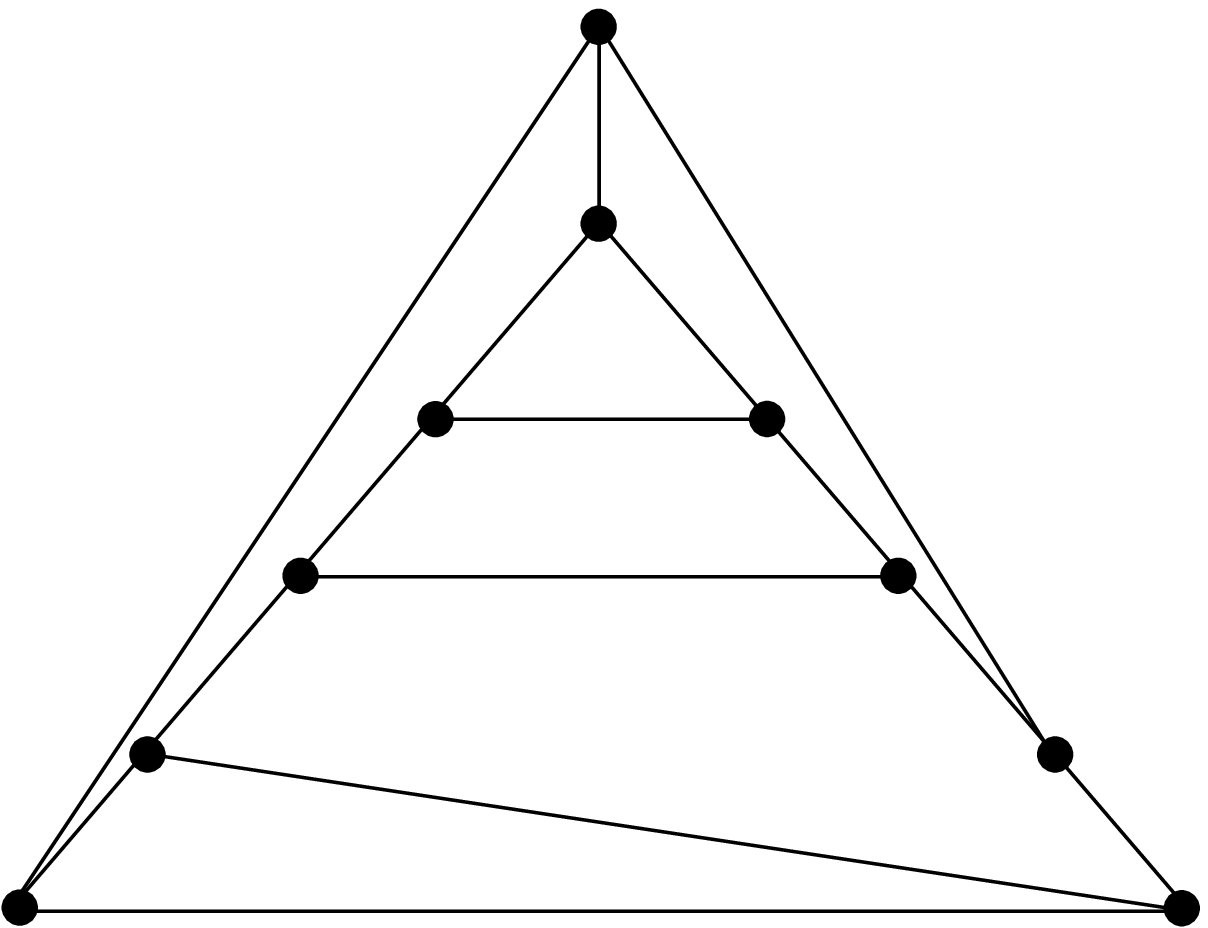} & $96$ &
      \includegraphics[width=60mm]{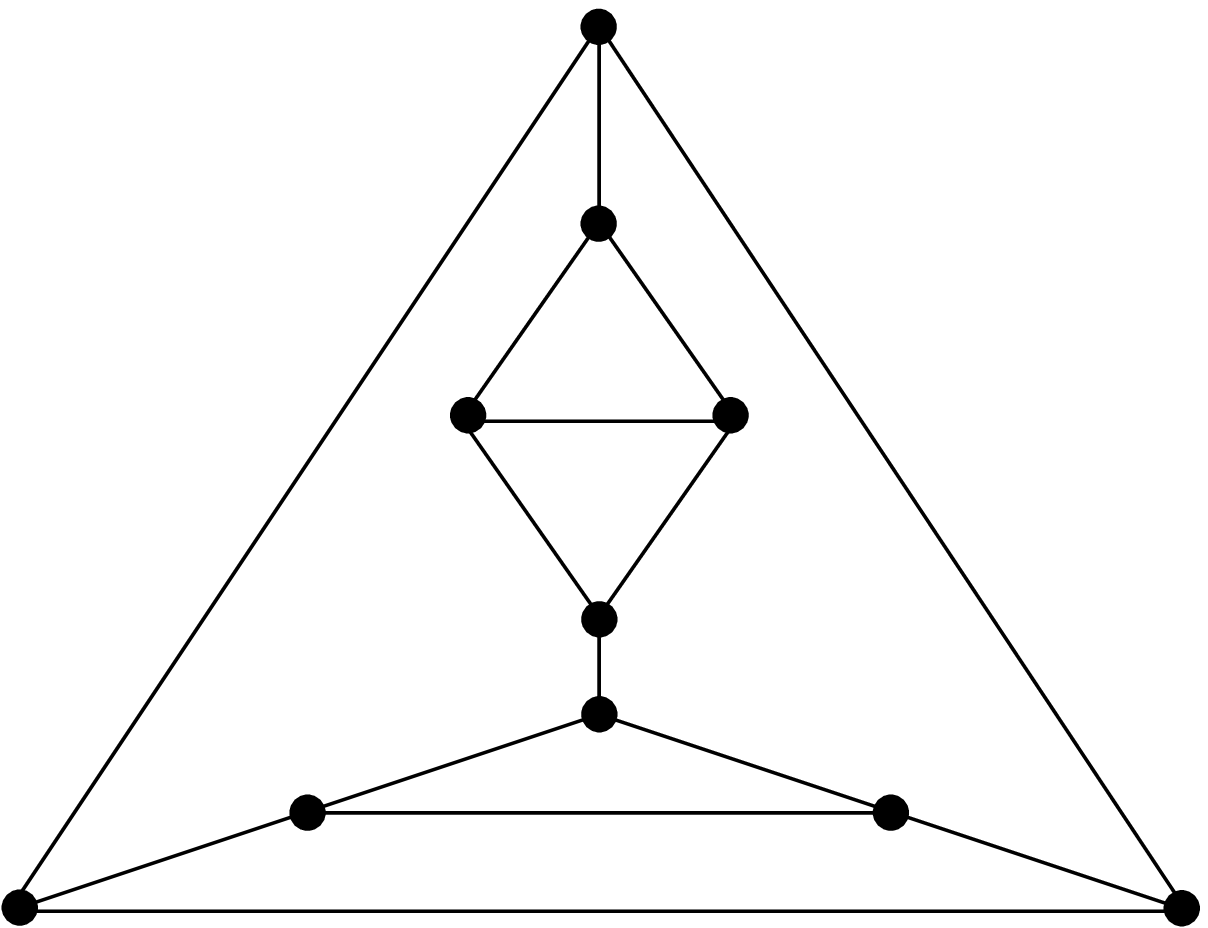} & $192$ \\
      \hline
       \includegraphics[width=60mm]{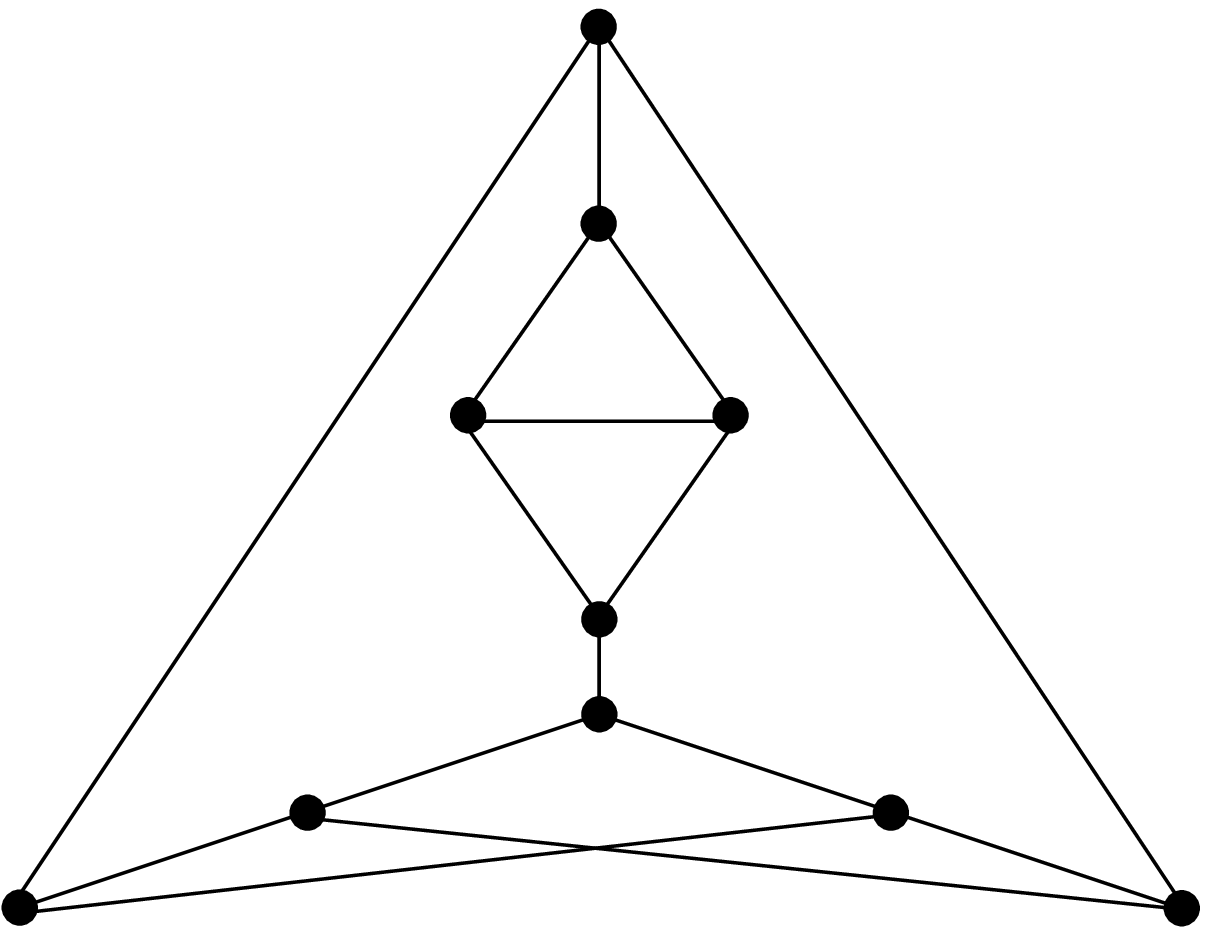} & $384$ &
      \includegraphics[width=60mm]{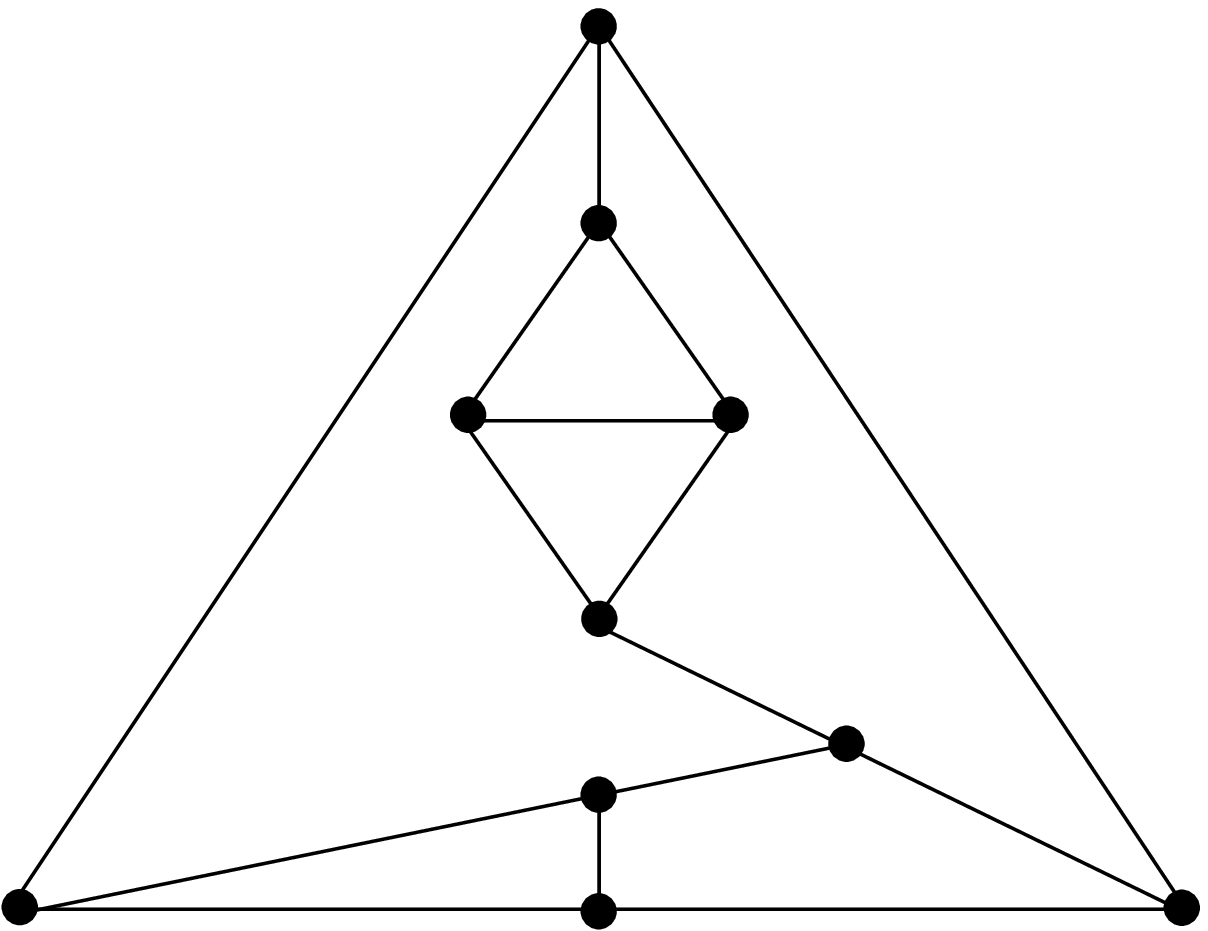} & $399$ \\
           \hline
              \includegraphics[width=60mm]{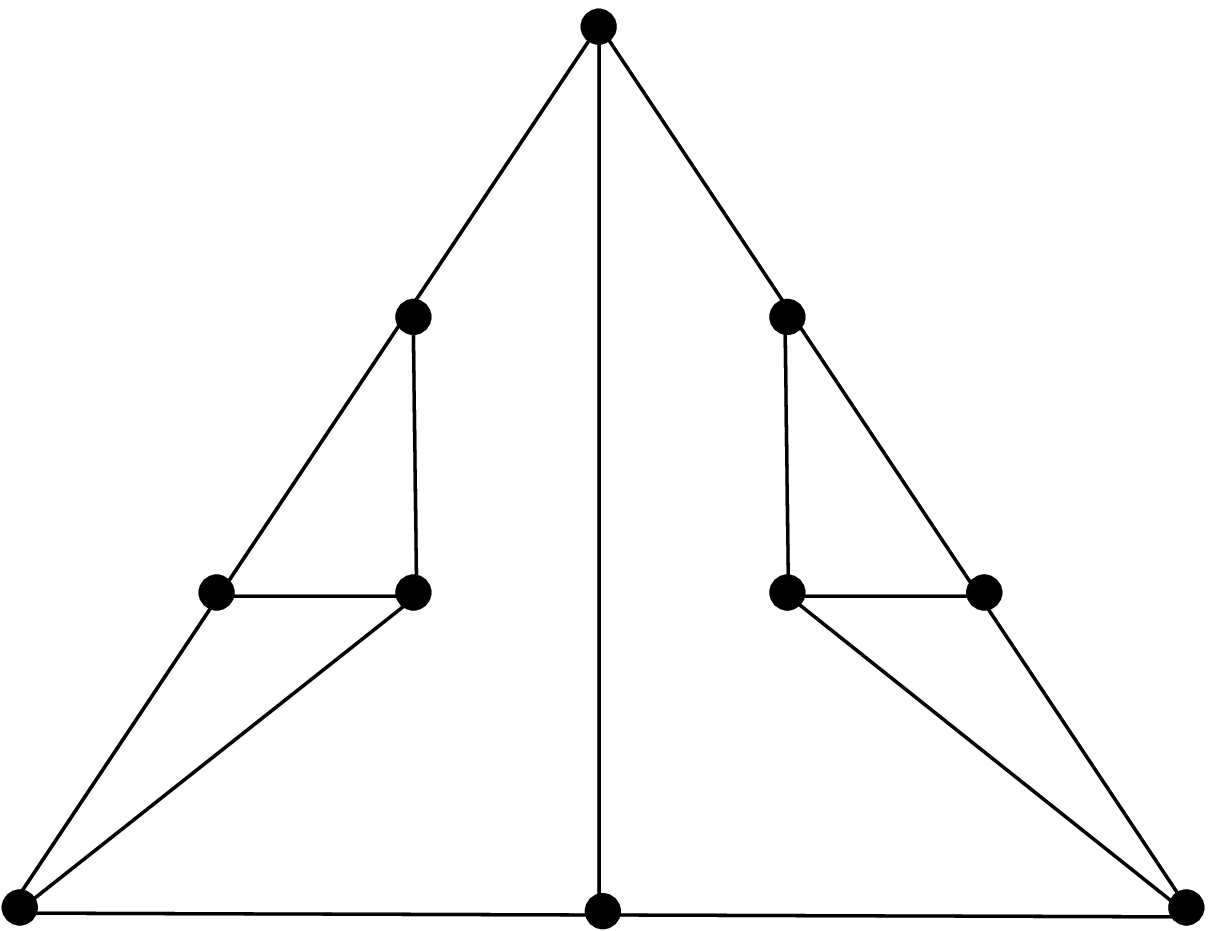} & $576$ &
      \includegraphics[width=60mm]{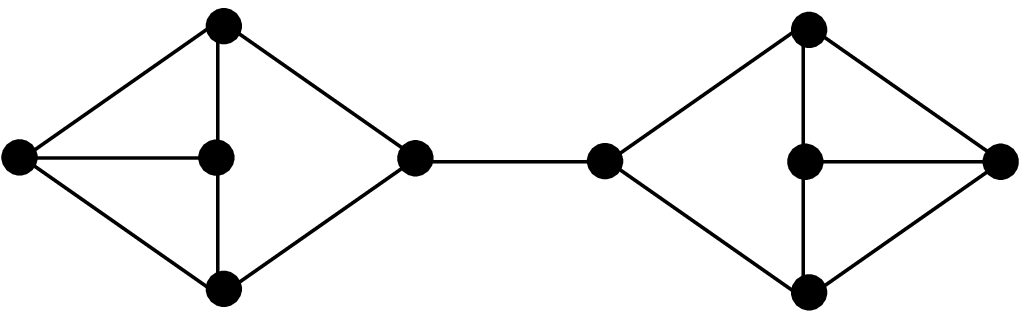} & $1152$\\
\hline
     \includegraphics[width=60mm]{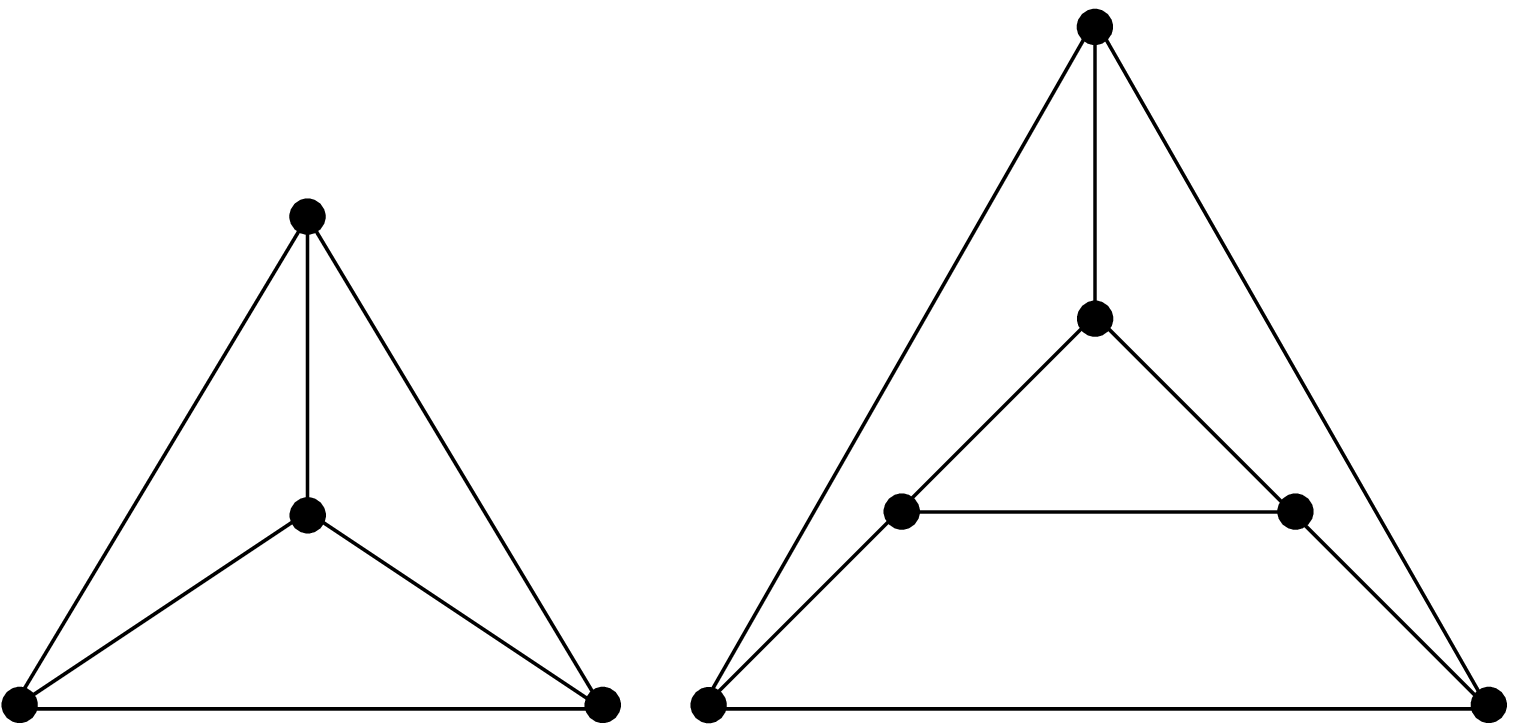} & $1728$ &
      \includegraphics[width=60mm]{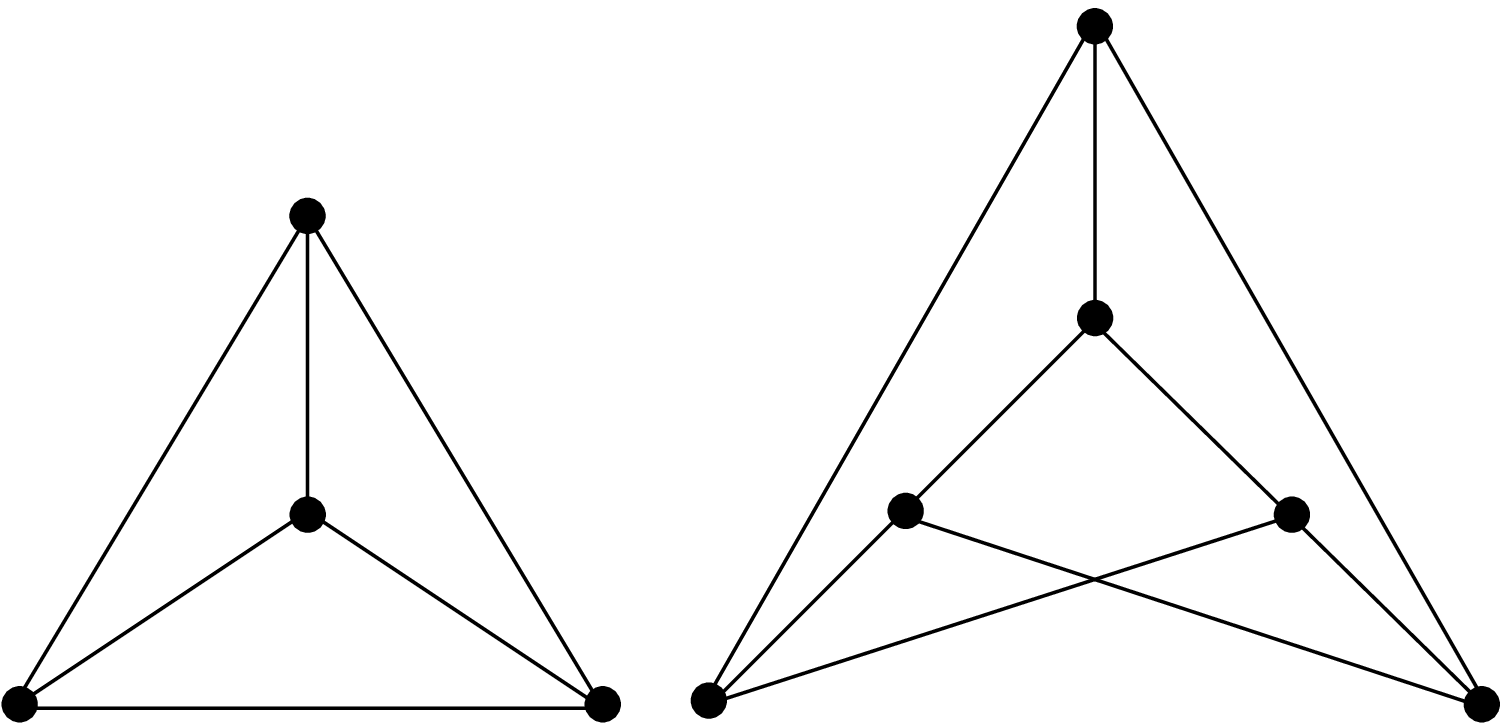} & $2160$  \\
             \hline

  \end{tabular}
         \caption{Rank $10$ Coxeter graphs with incidence index $\mc{I}=3$,  Coxeter exponents $2$ or $3$, and signature $(2 \big|_{+}, 8\big|_{-})$. None of these graphs correspond to line-incidence diagrams of geometric configurations of the type $(n_m,g_3)$.}
\end{table}
%\label{table:finite}
\end{center}

\begin{center}
\begin{table}
  \begin{tabular}{ |m{65mm}|m{10mm}|m{65mm}|m{10mm}|}
\hline
 Coxeter Graph $\mbb{C}_{\mf{C}}$ & $\det \hs C$ & Coxeter Graph  $\mbb{C}_{\mf{C}}$ & $\det\hs C$ \\
    \hline
     \hline
     \includegraphics[width=60mm]{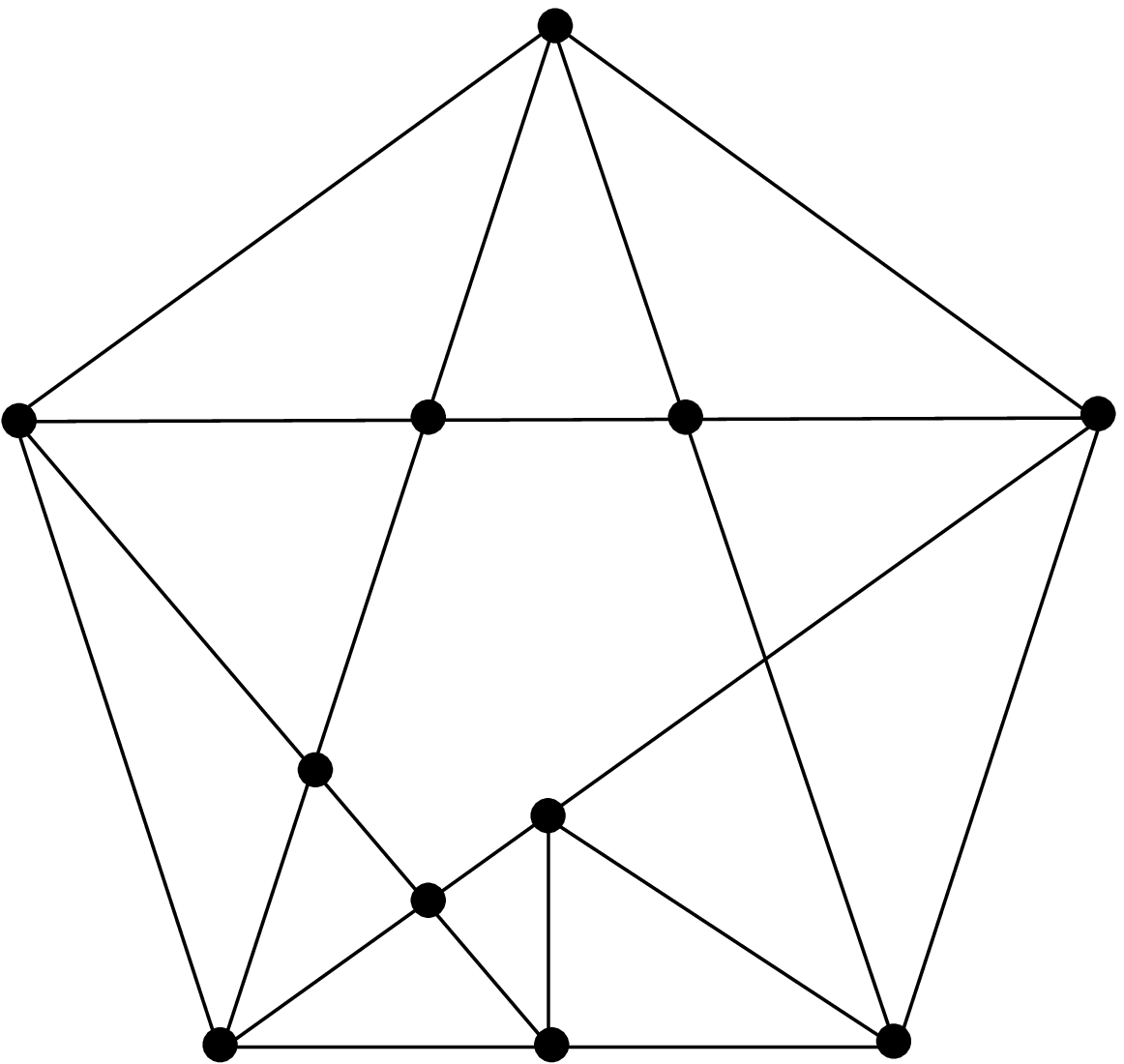} & $-3698$ &
      \includegraphics[width=60mm]{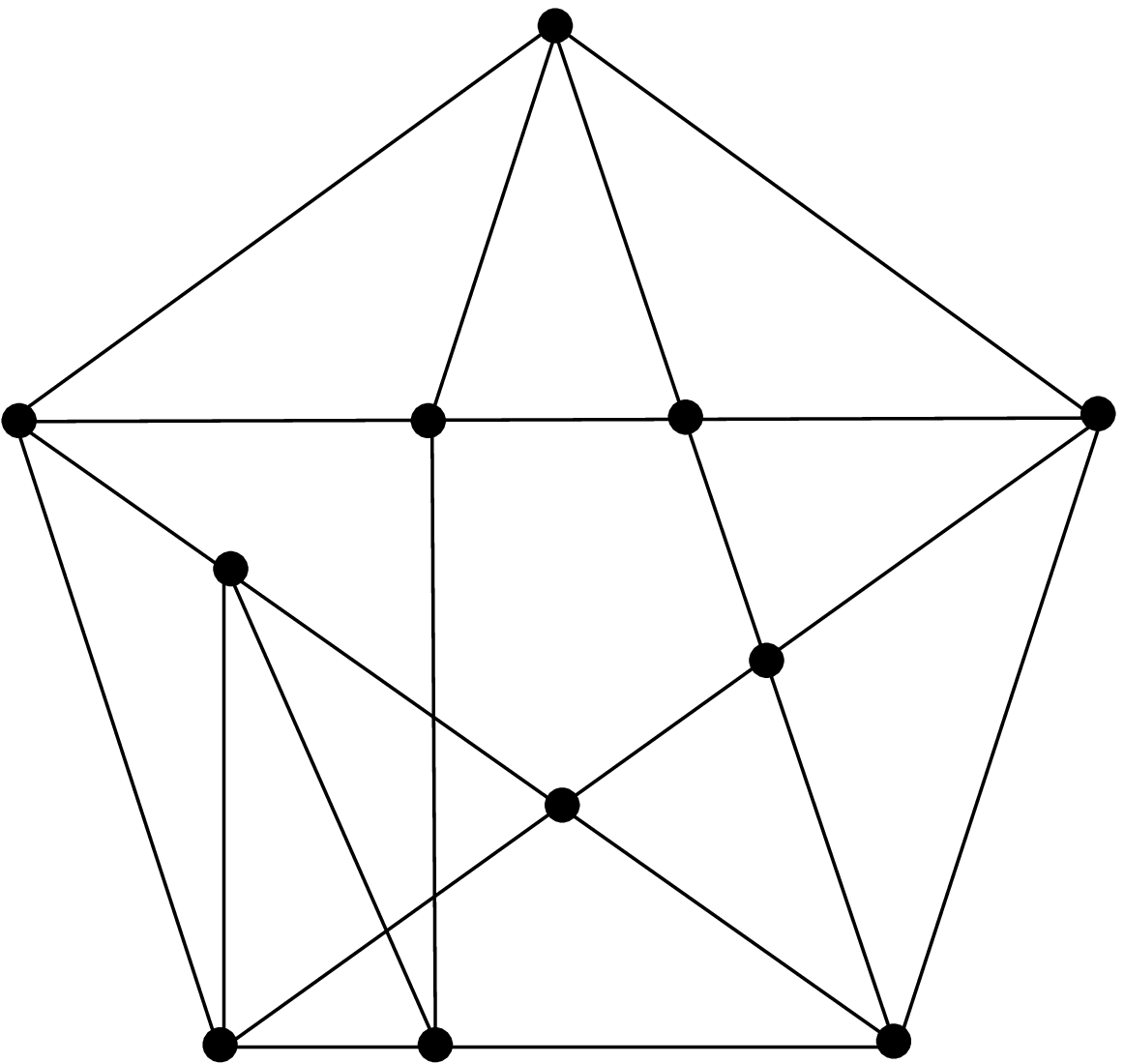} & $ -666 $  \\
    \hline
      \includegraphics[width=65mm]{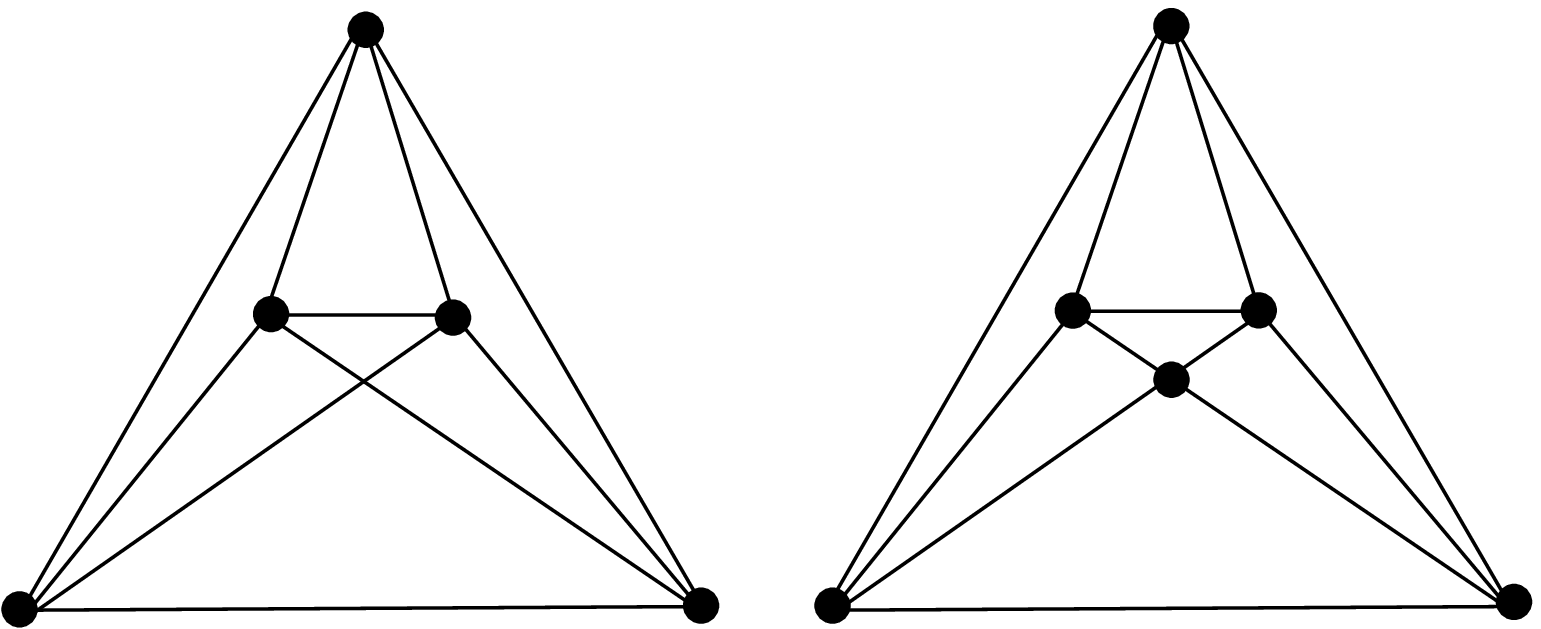} & $41472$ &
      \includegraphics[width=60mm]{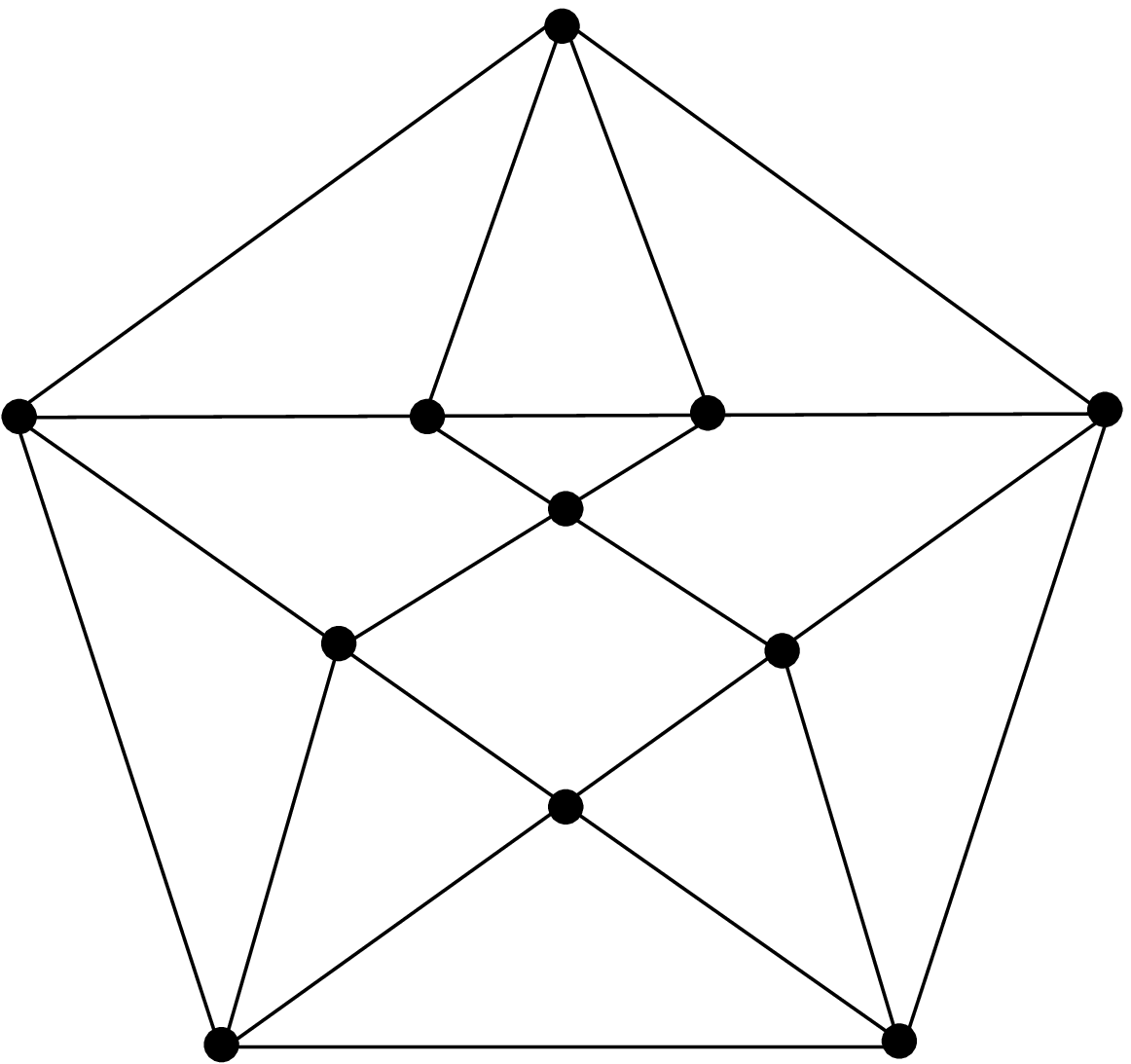} & $1248$ \\
      \hline

  \end{tabular}
         \caption{A selection of the Rank $11$ Coxeter graphs with incidence index $\mc{I}=4$ and Coxeter exponents equal to $2$ or $3$.  The top left graph is the one whose determinant is most negative and the bottom left is the one whose determinant is the most positive. None of these graphs correspond to line-incidence diagrams of geometric configurations of the type $(n_m,g_3)$.}
\end{table}
%\label{table:finite}
\end{center}

\strut\pagestyle{empty}
\newpage
\section*{Erratum}
In \cite{Henneaux:2006bw} we classified all rank $10$ (resp. $11$) Coxeter groups with incidence index $\mathcal{I}$  equal to $3$ (resp. $4$) and Coxeter exponents
equal to $2$ or $3$. This problem is equivalent to the classification of
all symmetric rank $10$ (resp. $11$) Cartan matrices with off-diagonal components 
equal to $0$, except for three (resp. four) on each line equal to $-1$. The method we used was to first construct a redundant set of all such Cartan matrices, after which we extracted from this the subset of matrices differing by their set of eigenvalues (characteristic polynomial). In this way we obtained the $19$ different rank $10$ Coxeter groups, but only $252$ of the $266$ existing rank $11$ Coxeter graphs with these properties \cite{web}. The origin of this discrepancy is due to the inadequacy of the method adopted. Indeed, there exist distinct symmetric rank 11 Cartan matrices that have the same set of eigenvalues, but are not equivalent, in the sense that they cannot be related through conjugation by any permutation matrix. 

\begin{table}[h]
\begin{center}
  \begin{tabular}{ |m{60mm}|m{60mm}|}
     \hline
     & \\
   \hspace{4mm}   \includegraphics[width=45mm]{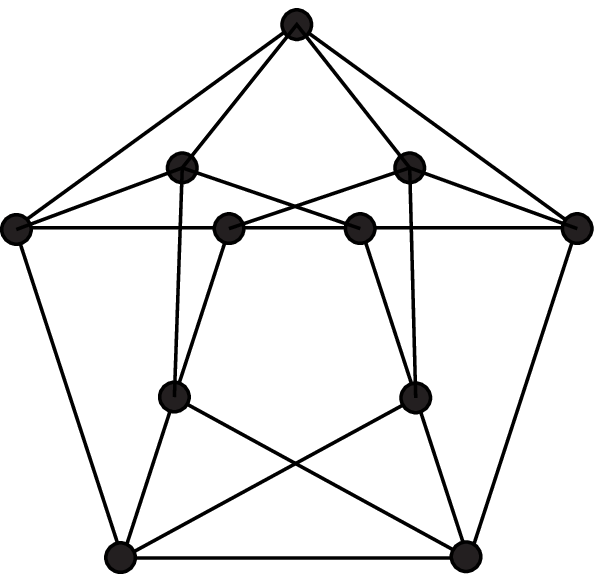} &
    \hspace{6mm}   \includegraphics[width=40mm]{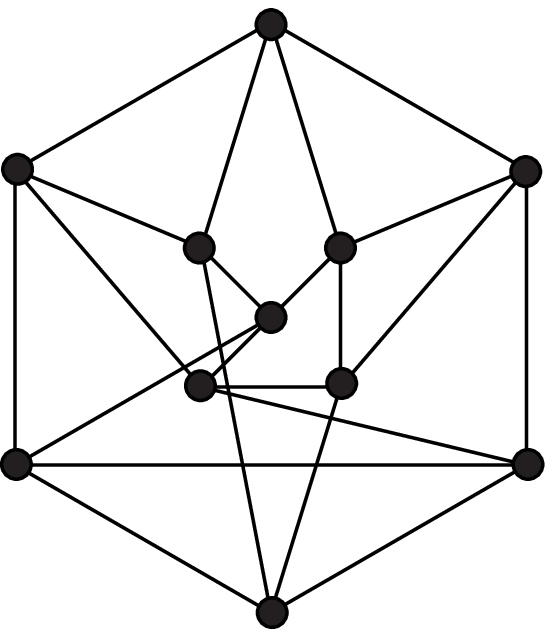}   \\
    & \\
      \hline
  \end{tabular}
         \caption{An example of two rank $11$ Coxeter graphs with incidence index $\mathcal{I}=4$. The left hand graph admits a $\mathbb Z_2$ automorphism group, while the right hand one does not admit any non-trivial automorphism. Both have the same characteristic polynomial: $-X^{11}+22 X^{10}-198 X^9+916 X^8-2123 X^7+1088
   X^6+6578 X^5-17658 X^4+19939 X^3-10988 X^2+2583
   X-170.$}
   \label{table:Errata}
\end{center}
\end{table}

Table \ref{table:Errata} provides an example of two such configurations, whose adjacency matrices give the same invariant polynomial, but nevertheless correspond to inequivalent Dynkin diagrams. Thus, given these considerations, the results presented in \cite{Henneaux:2006bw} have to be amended in the following way. 

\vspace{,3cm}

{\it There exist $266$ 
rank $11$ Coxeter groups that split into the following subsets:

\begin{itemize}
\item 73 Cartan matrices of Lorentzian signature, 15 of which
correspond to geometric configurations and so can be embedded into
$E_{11(11)}$. Among them two pairs have the same eigenvalues, but are inequivalent.
 \item 5 Cartan matrices with negative determinants,
all of which have signature $(3\big|_{-},8\big|_{+})$. \item 11
Cartan matrices with vanishing determinants, all of which have one
zero eigenvalue and one negative eigenvalue. Among them three inequivalent ones have the same set of eigenvalues, and seven can be
derived from geometric configurations. \item 1 Cartan matrix with
vanishing determinant and with two negative and one zero
eigenvalue. \item 176 Cartan matrices with positive determinants
but with signatures $(2\big|_{-},9\big|_{+})$. In terms of their set of eigenvalues they split into 157 singlets, 8 pairs and 1 triplet.
\end{itemize} }
All the 266 inequivalent Cartan matrices
are assembled in the file ``Coxeter11-4v4.nb'' which is included in
``Coxeterv4.zip'' that can be downloaded from the database arXiv.org
of Cornell University \cite{link}.

\vspace{.3cm}

\noindent \textbf{Acknowledgments:}  We are grateful to P-E. Caprace and H. M\'elot for discussions, and for providing us with useful references.

\end{document}